\documentclass[12pt,preprint]{aastex}
\usepackage{amssymb,amsmath,mathrsfs,color}
\usepackage{hyperref}
\newcommand{\arxiv}[1]{\href{http://arxiv.org/abs/#1}{arXiv:#1}}
\newcommand{\project}[1]{\textsl{#1}}
\newcommand{\An}{\project{Astrometry.net}}
\newcommand{\flickr}{\project{flickr}}
\newcommand{\yahoo}{Yahoo!}
\newcommand{\Yahoo}{\project{\yahoo}}
\newcommand{\YWS}{\project{\yahoo\ Web Search}}

\newcommand{\foreign}[1]{\emph{#1}}
\newcommand{\etal}{\foreign{et\,al.}}
\newcommand{\ie}{\foreign{i.e.}}
\newcommand{\etc}{\foreign{etc.}}
\newcommand{\figref}[1]{\figurename~\ref{#1}}
\newcommand{\Figref}[1]{\figref{#1}}
\newcommand{\thou}{,\!000}

\newcommand{\paramvector}[1]{\boldsymbol{#1}}
\newcommand{\pointing}{\paramvector{\alpha}}
\newcommand{\fovpars}{\paramvector{\Omega}}
\newcommand{\orbitpars}{\paramvector{\omega}}
\newcommand{\hyperpars}{\paramvector{\theta}}
\newcommand{\position}{\paramvector{x}}
\newcommand{\velocity}{\paramvector{v}}
\newcommand{\uniform}{\mathrm{uniform}}
\newcommand{\tmin}{t_\mathrm{min}}
\newcommand{\tmax}{t_\mathrm{max}}
\newcommand{\pgood}{p_\mathrm{good}}
\newcommand{\pempirical}{p_\mathrm{emp}}
\newcommand{\exif}{\mathrm{EXIF}}
\newcommand{\pexif}{p_\exif}
\newcommand{\texif}{t_\exif}
\newcommand{\pfg}{p_\mathrm{fg}}
\newcommand{\pbg}{p_\mathrm{bg}}

\newcommand{\unit}[1]{\mathrm{#1}}
\renewcommand{\day}{\unit{d}}
\renewcommand{\year}{\unit{yr}}
\newcommand{\AU}{\unit{AU}}

\newcounter{thumbnail}
\renewcommand{\thethumbnail}{\alph{thumbnail}}
\newcommand{\tlabel}[1]{\textsl{({#1})}}
\newcommand{\chumbnail}[4]{\includegraphics[width=#1\textwidth, height=#1\textwidth, keepaspectratio=true]{#2}\refstepcounter{thumbnail}\raisebox{1ex}{\makebox[0in][r]{\textcolor{green}{{#4}\tlabel{\thethumbnail}\rule[-1.5ex]{0.5ex}{0ex}}}}\label{#3}}
\newcommand{\shumbnail}[3]{\chumbnail{#1}{#2}{#3}{}}
\newcommand{\thumbnail}[3]{\chumbnail{#1}{#2}{#3}{$\ast$}}

\begin{document}

\title{Searching for comets on the World Wide Web:\\
       The orbit of 17P/Holmes from the behavior of photographers}
\author{Dustin~Lang\altaffilmark{1,2} \&
        David~W.~Hogg\altaffilmark{3,4}}
\altaffiltext{1}{Princeton University Observatory, Princeton, NJ, 08544, USA}
\altaffiltext{2}{to whom correspondence should be addressed:
                 dstn@cmu.edu}
\altaffiltext{3}{Center for Cosmology and Particle Physics,
                 Department of Physics, New York University,
                 4 Washington Place, New York, NY, 10003, USA}
\altaffiltext{4}{Max-Planck-Institut f\"ur Astronomie,
                 K\"onigstuhl 17, D-69117, Heidelberg, Germany}

\begin{abstract}
We performed an image search for ``Comet Holmes,'' using the
\Yahoo\ Web search engine, on 2010~April~1.  Thousands of images were
returned.  We astrometrically calibrated---and therefore vetted---the
images using the \project{Astrometry.net} system.  The calibrated
image pointings form a set of data points to which we can fit a
test-particle orbit in the Solar System, marginalizing over image
dates and detecting outliers.  The approach is Bayesian and the model
is, in essence, a model of how comet astrophotographers point their
instruments.  In this work, we do not measure the position of the
comet within each image, but rather use the celestial position of the
whole image to infer the orbit.  We find very strong probabilistic
constraints on the orbit, although slightly off the JPL ephemeris,
probably due to limitations of our model.  Hyperparameters of the
model constrain the reliability of date meta-data and where in the
image astrophotographers place the comet; we find that $\sim
70$~percent of the meta-data are correct and that the comet typically
appears in the central third of the image footprint.  This project
demonstrates that discoveries and measurements can be made using data
of extreme heterogeneity and unknown provenance.  As the size and
diversity of astronomical data sets continues to grow, approaches like
ours will become more essential.  This project also demonstrates that
the Web is an enormous repository of astronomical information; and
that if an object has been given a name and photographed thousands of
times by observers who post their images on the Web, we can
(re-)discover it and infer its dynamical properties.
\end{abstract}


\section{Introduction}

The Web bristles with billions of images: on Web pages, in public
photo-sharing sites, on social networks, and in private email and
file-sharing conversations.\footnote{The \emph{Flickr Blog} reported
  their 5 billionth image upload on 2010-09-19
  (http://blog.flickr.net/en/2010/09/19).} A tiny fraction but
\emph{enormous number} of these images are \emph{astronomical}
images---images of the night sky in which astronomical sources are
visible.  This is true even if we exclude from consideration
scientific collections such as those of professional observatories and
surveys and only count the images of hobbyists, amateurs, and
sight-seers.  In principle these images, taken together, contain an
enormous amount of information about the astronomical sky.  Of course
they have no scientifically responsible provenance, have never been
``calibrated'' in any sense of that word, and were (mainly) taken for
purposes that are not at all scientific.  But having been generated
from CCD-like measurements of the intensity field, they cannot help
but contain important scientific information.  The Web is, therefore,
an enormous and virtually unexploited sky survey.

It is difficult to estimate the total number of astronomical images on
the Web, and even harder to estimate the total data throughput
(\foreign{\'etendue} or equivalent measure of scientific information
content).  However, by any estimate, it is extremely large.  For
example, image search results for common astronomical subjects include
thousands of astronomical images.  The \flickr\ photo-sharing Web site
has an astrometry group (administered by the \An\ collaboration; more
below) with more than $23\thou$ photos, and its astronomy and
astrophotography groups have more than $35\thou$ and $45\thou$
respectively.  A search for the Orion Nebula on \flickr\ returns more
than 9000 images, which jointly contain significant information on
very faint stars and nebular features.  These numbers---derived solely
from \flickr\ searches---represent only a tiny fraction of the
relevant Web content.  Of course all these search results contain many
non-astronomical images, diagrams, fake data, and duplicates, so use
of them for science is non-trivial.

The technical obstacles to making use of Web data are immense: If
anything has been learned from our interaction with electronic
communication, it is that publisher-supplied or provider-supplied
meta-data about Web content are consistently missing, misleading, in
error, or obscure.  Indeed, when it comes to the astronomical
properties of imaging discovered on the Web, most providers do not
even know what we want in terms of ``meta-data''; we want calibration
parameters relating to image date, astrometric coordinate system,
photometric sensitivity, and point-spread function, and we want it in
machine-readable form.  The Virtual Astronomy Multimedia Project
(VAMP, \citealt{vamp}) has defined a format for placing astrometric
meta-data in image headers, but the goal of the project is to make
``pretty pictures'' searchable for education and public outreach
purposes, rather than science.  They do not consider the problem of
producing or verifying calibration meta-data; they assume correct
meta-data are provided along with the science images that are used to
produce the pretty pictures.  Even the Virtual Observatory
(\url{http://ivoa.net}), which concentrates on astronomical meta-data,
has no plan for ensuring that meta-data are \emph{correct}, and has no
machine-readable form for many quantities of great interest (such as
the detailed point-spread-function model); we cannot expect the
world's amateur astrophotographers to be better organized.

Two important changes are occurring in astronomy that are opening up
the possibility that we might exploit data collections as radically
confusing as that of the entire Web.  The first is that tools are
beginning to appear that can perform completely hands-free data
analysis tasks.  The best example so far is the
\An\ system, which can take astronomical imaging of
completely unknown provenance, and calibrate it astrometrically using
the data in the image pixels alone (\citealt{An}).

The second change is that there has been an enormous increase in the
amount and diversity of publicly available professional data---that
is, calibrated, trustworthy, science-oriented data in observatory,
sky-survey, and individual-investigator collections.  These
collections are so large and diverse that automated data analysis
tools that can trivially interact with extremely heterogeneous data
are necessary in many scientific domains.  That is, much of the
technology required for exploitation of the Web-as-sky-survey is
required for \emph{any} mature, data-intensive scientific
investigation.

We have been exploring some of these ideas with the \An\ project.  Not
only has the system calibrated thousands of images taken by amateurs
and hobbyists, we have interfaced the system with \flickr\ (Stumm
\etal, forthcoming).  When a user adds an image to the ``astrometry''
group, an automated ``bot'' downloads the image, calibrates it with \An,
and then posts machine-readable calibration results to the image's
page on \flickr; these have been dubbed ``astro-tags''.  The bot also
adds annotations to the image, marking named stars and galaxies from
the Messier and NGC/IC catalogs.  We make use of the
\flickr\ Application Programming Interface (API); many image and
data-sharing sites offer APIs that allow scriptable access to the data
they hold.  The success of our \flickr\ bot suggests that automated
maintenance of a heterogeneous crowd-sourced sky survey might be
possible in the future.

In this paper, we explore some of the ideas around the
Web-as-sky-survey, by performing a scientific investigation of Comet
17P/Holmes using Web-discovered, human-viewable (JPEG) images alone.
Although we do in principle learn things about Comet Holmes, our main
interest is in developing and testing new technologies for
observational astrophysics.  This project leverages the tendency of
humans to point their cameras and telescopes towards interesting
things, the ability of \Yahoo\ (or any other search engine) to
classify and organize their images, and the ability of \An\ to figure
out after the fact where they were pointing.  What we do is related to
other citizen-science projects, like the \project{GalaxyZoo}
(\citealt{galaxyzoo}) or the monitoring projects of the
\project{AAVSO} (\url{http://aavso.org}), except that the participants
here are entirely unwitting.  We end up showing that there is
substantial scientific content in the data taken by citizen observers,
even if they have not committed to particular scientific goals; we
also show that it is possible to extract scientific information from
observations of which the provenance is unknown.

\section{Data and calibration}

Our data collection began with a search of the World Wide Web.  We
used the \mbox{\project{pYsearch}} (\citealt{pysearch}) code to access
the \YWS\ service.\footnote{%
  http://developer.yahoo.com/search/web/webSearch.html} On
2010~April~1, we searched for JPEG-format images using the query
phrase ``Comet Holmes.''  Due to a dramatic brightening during its
2007 apparition, Comet Holmes became a very popular and accessible
target for astrophotographers, so many images are available on the
Web.  Our search yielded approximately $10\thou$ total results, but
the \YWS\ API allowed only $1000$ results to be retrieved per query.
In order to broaden the result set, we performed an additional set of
searches: For each Web site containing an image in the original set of
results, we performed a query that was limited to that Web site.
These queries, performed later on 2010~April~16, produced an
additional $2741$ results (including some duplicates), for a total of
$2476$ unique results.  See \figref{fig:examples} for some example
images.

Next, we retrieved the images on 2010~April~16.  This yielded a total
of $2309$ valid JPEG images.  After removing byte-identical images,
$2241$ unique images remained.  We then ran the \An\ code on each
image to perform astrometric calibration.  $1299$ images were
recognized as images of the night sky and astrometrically calibrated.
These images form the data set we use in our analysis below.
\figref{fig:footprints} shows the footprints of the images on the sky.

Many of the images in this data set are annotated images or diagrams
such as finding charts or illustrations of the comet's orbit.  Some of
these diagrams were recognized by \An\ as images of the sky.  This can
be seen in the co-added image in \figref{fig:footprints}, where there
are clearly lines connecting the stars that form the constellation
Perseus.

Of the $1299$ images in our data set, $422$ have timestamps in the
image headers (``Exchangeable image file format'' or EXIF headers).
The distribution of timestamps is shown in \figref{fig:imgstats}.  On
2007 Oct 24, Comet 17P/Holmes brightened by more than 10~mag
(\citealt{outburst}), generating considerable public interest and
making it a very popular and accessible observing target in the
amateur astronomy community.  The distribution of image timestamps
shows a large spike at this time.

We evaluate the accuracy of the image timestamps by asking, for each
image, whether the comet would appear within the celestial-coordinate
bounds of the image at its stamped time.  We find that the majority of
the timestamps are consistent, and that inconsistent timestamps are
typically late rather than early; we assume this is because some of
the images have been post-processed and the timestamp represents the time the
image was last edited rather than the time the image was taken.  See
\figref{fig:exiftimes}.

\figref{fig:imgstats} shows the distribution of angular scales
of the images in our data set.  The distribution peaks around $3$
square degrees.  Also shown is the distribution of exposure times
reported in the EXIF headers.

\section{Orbit inference}

We take the approach of generative modeling; that is, we construct a
well-defined approximation to the probability of the data given the
model.  We take the ``data'' to be the pointing (on the sky) of each
astrometrically calibrated image (as determined by \An); recall that
the goal is to use the \emph{behavior} of astrophotographers (in
pointing their cameras) to find the gravitational orbits of objects in
the sky.  We treat the time at which each image is taken as a hidden
``nuisance'' parameter.

For each image $i$ there is a pointing $\pointing_i$ (two-dimensional
position or celestial coordinates on the sky).  These are the
\emph{data}.  The image was taken at time $t_i$, and has image
parameters $\fovpars_i$ (camera plate scale, image size, orientation,
and reported EXIF timestamp if there is one), taken to be known.  In
addition, the comet has orbital parameters $\orbitpars$, which can be
thought of as semi-major axis, eccentricity, inclination, longitudes,
\etc, or equivalently a 3-dimensional position $\position$ and
velocity $\velocity$ at a chosen epoch.  We choose the latter for
inference simplicity, and use as the epoch JD $2454416.0$
(2007~Nov~12).  Finally, there are three additional nuisance
\emph{hyperparameters} $\hyperpars$ that will appear as we go.  In
this work we consider only the 2007 apparition of the comet.  Our data
set does include at least one image of the comet during an earlier
apparition (1892; see \figref{fig:examples}\tlabel{\ref{barnard}}),
but we chose not to attempt to fit multiple apparitions.

The single-image likelihood
$p(\pointing_i|\fovpars_i,\orbitpars,\hyperpars)$ is a marginalization
over time $t_i$ of the time-dependent single-image likelihood,
$p(\pointing_i|t_i,\fovpars_i,\orbitpars,\hyperpars)$:
\begin{eqnarray}\displaystyle
p(\pointing_i|\fovpars_i,\orbitpars,\hyperpars) &=& \int
p(\pointing_i|t_i,\fovpars_i,\orbitpars,\hyperpars)
\,p(t_i|\fovpars_i,\hyperpars)\,dt_i \quad , 
\end{eqnarray}
where the time-dependent single-image likelihood is a mixture of
``foreground'' (inlier) and ``background'' (outlier) components:
\begin{eqnarray}\displaystyle
p(\pointing_i|t_i,\fovpars_i,\orbitpars,\hyperpars)
  &=& \pgood\,\pfg(\pointing_i|t_i,\fovpars_i,\orbitpars,\hyperpars)
    + [1-\pgood]\,\pbg(\pointing_i) \quad .
\end{eqnarray}
The components are:
\begin{eqnarray}\displaystyle
\pfg(\pointing_i|t_i,\fovpars_i,\orbitpars,\hyperpars)
  &=& \left\{\begin{array}{ll}
      [\eta\,\Omega_i]^{-1} & \mbox{comet in $\eta$ sub-image} \\
                          0 & \mbox{comet not in $\eta$ sub-image}
    \end{array}\right.
\\
\pbg(\pointing_i)
  &=& [4\pi]^{-1}
\quad .
\end{eqnarray}
The hyperparameters $\hyperpars$ include $\pgood$, $\eta$, and
$\pexif$ (discussed below).  Here, $\pgood$ is the probability that
the image really \emph{is} a picture intentionally taken of (generated
by) the comet, $\pfg(\cdot)$ is a ``foreground'' model, which gives
high likelihood when the comet (with orbital parameters $\orbitpars$
at time $t_i$) is inside the image, and $\pbg(\cdot)$ is a
``background'' model, with no dependence on the comet or time, that
describes images that are in our data set but do not contain the comet
(perhaps because they were incorrectly returned by the Web search
engine).  The $4\pi$ in $\pbg(\cdot)$ is the solid angle of the whole
sky.  The hyperparameter $\eta$ (subject to $0<\eta<1$) controls the
fractional size of the central region of an image in which
astrophotographers place comet subjects, $\Omega_i$ is the solid angle
covered by image $i$, and the ``$\eta$ sub-image'' is the central
$\eta$ of the image.  In detail, we define the $\eta$ sub-image to
have the same aspect ratio as the whole image, centered at the same
point, but smaller in angular size by $\sqrt{\eta}$ along both
dimensions.

Our Bayesian prior probability distribution function (PDF) over time,
$p(t_i|\fovpars_i,\hyperpars)$, turns out to be crucial to good
inference in this problem, in part because trivial or wrongly
uninformative time PDFs lead to highly biased answers, a point to
which we will return below.  We expect that a large fraction of the
image EXIF timestamps (where they exist) are correct, but at the same
time we cannot trust them completely.  We construct an empirical
``cheater'' prior $\pempirical(t)$ based on the empirical histogram of
extant EXIF timestamps as follows: We construct a grid of
non-overlapping bins in time of width 8~d between $\tmin=$2007~July~1
and $\tmax=$2008~May~1.  We count EXIF timestamps in these bins, and
then add 1 to every bin (so no bins have counts of zero).  We then
normalize so that the integral of $\pempirical(t)$ is unity.  This
empirical prior is shown in \figref{fig:empirical}.  Given any image
$i$ with image parameters $\fovpars_i$, the PDF for time $t_i$ is
\begin{eqnarray}\displaystyle
p(t_i|\fovpars_i,\hyperpars)
  &=& \left\{\begin{array}{ll}
      \pempirical(t_i)
        & \mbox{if no $\texif$ in $\fovpars_i$} \\
      \pexif\,p(t_i|\texif) + [1-\pexif]\,\pempirical(t_i)
        & \mbox{if $\texif$ in $\fovpars_i$} \\
    \end{array}\right.
\nonumber\\
p(t_i|\texif)
  &=&\uniform(t_i|\texif-[0.5~\day],\,\texif+[0.5~\day])
\quad ,
\end{eqnarray}
where $\pexif$ is the third hyperparameter in $\hyperpars$ and the
probability that a given EXIF timestamp is reliable, $\uniform(x|A,B)$
is the top-hat or uniform PDF for $x$ between $A$ and $B$, $\texif$ is
the reported EXIF timestamp, and we have subtracted and added
$0.5~\day$ because the EXIF format contains no time zone information
and this is the span of possible time zones.  In short, if an image
does not contain an EXIF timestamp, the model uses the empirical
distribution of time stamps.  If an image does contain an EXIF
timestamp, it is likely to be correct so the model assigns fraction
$\pexif$ of the probability mass to a 24-hour window around the
timestamp, but also hedges by including fraction $1-\pexif$ of the
empirical distribution.  An example is shown in
\figref{fig:empirical}.

The single-image likelihood
$p(\pointing_i|\fovpars_i,\orbitpars,\hyperpars)$ is the probability
that an image would be taken at coordinates $\pointing_i$ given the
camera properties $\fovpars_i$, the comet's trajectory $\orbitpars$,
and our model hyperparameters $\hyperpars$, integrated (marginalized)
over the time period we consider.  We do not specify the exact time
$t_i$ of each image; we instead specify a probability distribution of
times and integrate over it (and this integration \emph{requires} us
to be Bayesian).
In order to evaluate this likelihood, we compute the comet trajectory
on a fine time grid and perform the time integral numerically as a sum
over grid points.  For dynamical integration we use a Keplerian two-body
celestial mechanics code implemented in Python by
\An\ for both the comet and the Earth--Moon barycenter (EMB); we take
the initial conditions of the EMB from JD $2454101.5$ (2007~Jan~1).
For simplicity we take the EMB to be the observer's location, thus
ignoring the effect of parallax.  At the precision of the data, the
finite light-travel time in the Solar System is significant; we
include it when we consider the observed position of the comet as a
function of time.  For the numerical integrals, we simply convert
observed Solar-system directions to positions on the celestial sphere,
and positions on the celestial sphere to image positions (to determine
whether particular comet instances are inside particular images) with
the \An~world-coordinate system libraries (\citealt{An}).
Outliers---images that do not contain the comet---will only by chance
intersect the comet trajectory so will have low likelihood under the
foreground model $\pfg(\cdot)$; the background model $\pbg(\cdot)$,
which has no dependence on the orbital parameters, will dominate the
likelihood.  As with the time $t_i$, we need not explicitly estimate
whether any given image is an outlier; we instead model the data as a
mixture of the foreground distribution (``inliers'') and a background
distribution (``outliers''), and sum (marginalize) over the two
possibilities.

The total likelihood is the product of the individual-image
marginalized likelihoods, and the posterior PDF for the parameters
$\orbitpars$ and $\hyperpars$ is proportional to the total likelihood
times a prior.  We take this prior to be Gaussian in comet position
$\position$ with three-dimensional isotropic Gaussian variance of
$[1~\AU]^2$, a beta distribution in squared velocity
$v^2\equiv\velocity\cdot\velocity$ between $v^2=0$ and the $v^2$ that
just unbinds the comet, with beta-distribution parameters $\alpha=1$
and $\beta=3$.  We take the prior to be flat in the range 0 to 1 for
the probability hyperparameters $\pgood$ and $\pexif$ and (improperly)
flat in $\ln(\eta)$ for the fractional hyperparameter $\eta$.  These 9
parameters (three position components, three velocity components, two
probabilities, and one fraction) are the parameters in which we
perform our Markov Chain Monte Carlo (MCMC) sampling.  We perform the
sampling with a Python implementation (\citealt{emcee})
of an affine-invariant ensemble sampler
(\citealt{sampler}, \citealt{hou}) using an ensemble of 64
walkers and Python multiprocessor support.

We initialize the MCMC using the following heuristic.  Since we know
we have many images taken during a short window around the comet's
outburst, we select images with timestamps within a week of the median
date.  Of these images, we keep those whose centers are within 5
degrees of the median right ascension and declination.  This gives us
a set of $136$ images to which we fit lines for right ascension and
declination with respect to time, weighting by the inverse image
extents.  Assuming the comet is $1~\AU$ away from Earth and moving
perpendicular to the line of sight, we convert the projected position
to a three-dimensional position $\position_0$ and velocity
$\velocity_0$ at the epoch.  We choose eyeballed-sensible initial
values for the hyperparameters: ${\pgood}_0 = 0.85$, ${\pexif}_0 =
0.75$, and ${\eta}_0 = 0.4$.  To initialize the ensemble of MCMC
samples, we add Gaussian noise with standard deviation $10^{-5}~\AU$
to the position $\position_0$, $10^{-3}~\AU/\year$ to the
velocity $\velocity_0$, and $10^{-2}$ to each of ${\pgood}_0$,
${\pexif}_0$, and ${\eta}_0$.  The details of the initialization are
not critical; the MCMC sampler will explore the parameter space and
find a good solution in a reasonable number of iterations as long as
we initialize it somewhat near the solution.


There are several substantial limitations to this model: The prior
does not even come close to representing our true prior knowledge
about comets, particularly ones that are observed by photographers and
posted to the Web.  The total likelihood (being a product of
individual-image likelihoods) assumes all the data are independent,
but in reality some of the images found by the Web search are repeats,
duplicates, or derived images from others.  Most importantly, we make
no attempt to \emph{find the comet in the image}.  This is a model of
how astronomers point their cameras, not of the visible comet itself.

The results of the inference are shown in \figref{fig:traj} as a set
of sample trajectories drawn from the Markov chain.  These samples are
effectively drawn from the posterior PDF marginalized over the
hyperparameters.  The small dispersion among the samples show that the
data---just the pointings of a set of heterogeneous images---are
incredibly informative about the comet orbit.  In
\figref{fig:param-hists} we show our estimates for the standard
orbital elements, along with the values from the \project{Horizons}
system from the Jet Propulsion Lab (JPL; \citealt{jpl}), which we take
to be authoritative.  Most of our parameter estimates are a few
standard deviations away from the JPL values.  \Figref{fig:three-d}
shows that the three-dimensional orbit we infer is quite close to the
JPL orbit in the regions where we have data, and the variance in our
inferred orbit correctly increases in the regions where we lack data.

\section{Discussion}

We have shown that if a Solar System body has been \emph{named} and
hundreds of astrophotographers around the world have deliberately
\emph{photographed it}, we can recover its dynamical properties by a
Web search operation followed by a large amount of computation.  All
the inference is done on image positions; we never look at the content
of the images at all.  This effectively makes the model a model of
astrophotographers, because the image pointings are a record of where
human observers pointed their telescopes and cameras.  The six
dynamical parameters are parameters of the comet to be sure, but the
three hyperparameters are affected by human behavior as well as by
properties of the comet.  The probability $\pgood$ relates to the
purity of image search on the Web, the probability $\pexif$ relates to
the reliability of astrophotographers' Web-published image meta-data,
and the fraction $\eta$ relates to how astrophotographers frame their
images.
We find hyperparameter $\pgood \sim 0.9$, which indicates that a large
majority of the images returned by the Web search and recognized by
\An\ do indeed contain Comet Holmes.  We find $\pexif \sim 0.7$,
indicating that when images have timestamps in their headers, they are
often correct.  We find $\eta \sim 0.36$, indicating that
astrophotographers tend to place Comet Holmes in the middle third of
the image area.  

We expect that these hyperparameters will vary for different Web
search engines, query phrases, and comets.  Comets with distinctive
names are likely to be better indexed by search engines, faint comets
are likely to be photographed by different populations of observers,
and comets with long tails are likely to be framed differently in
photographs.  We expect that searching for the phrase ``17P/Holmes''
rather than ``Comet Holmes,'' for example, might result in images from
more technical astronomy Web sites rather than the popular press.  As
a result, we might find that a higher fraction of the images contain
the comet, leading to a higher inferred $\pgood$; we might find that
the images have different distributions of angular extents and
different framing, resulting in different $\eta$; and we might expect
that more of the images were captured with telescope-mounted CCDs,
saved in non-JPEG formats and subsequently processed to produce the
pretty picture, perhaps resulting in different timestamp properties
and therefore $\pexif$.


If we omit $\eta$ from our model (\ie, assume it is unity), we
effectively ignore the fact that astrophotographers frame their
subjects, and this in turn increases the positional uncertainty of the
comet, which results in poorer constraints on the orbit of the comet.
By including $\eta$, we give our model the freedom to learn how
astrophotographers compose their images.  The model prefers to set
$\eta$ to a rather small fraction, indicating that this freedom is
useful.  Including $\eta$ means that if the comet appears in the
corner of an image, the image will be treated as an outlier and will
not contribute to constraining the orbit.  However, this loss of
constraining power is balanced by stronger constraints from all images
in which the comet appears near the center.  In
\figref{fig:param-hists}, there are three peaks in the histogram of
$\eta$ values.  We assume the two sub-dominant peaks correspond to
$\eta$ values that omit one image or include one additional image in
the solution, relative to the dominant peak.

The model is exceedingly crude, and the fact that our results are
biased (the samples in \figref{fig:traj} are offset from the JPL
trajectory) is probably in part related to this crudeness.  The
centering model is extremely crude; in reality there is a distribution
of astrophotographers' behavior that it ought to describe.  The time
model involves a hard-set empirical prior that is not justified and
ought to be simultaneously optimized and marginalized out in the
inference (this would be a form of hierarchical inference like in
\citealt{hierarchical}).  The time-zone model (flat across all time
zones) is also not realistic, since some time zones are much more
populated with photographers than others.  Along those same lines,
there is an enormous amount of external information (weather data and
visibility calculations) that could further constrain the possible
times and time zones.  As with time zones, we do not attempt to model
the positions of the astrophotographers relative to the Earth-Moon
barycenter.  By ignoring the resulting parallax, we incur errors of
about $5$ arcsecond in our estimate of the comet's position.

Another crudeness is in the assumption of independent and identically
distributed draws in the likelihood.  This is not true in that some of
the Web images we find are crops, edits, or diagrams made from other
Web images.  That is, each image is not guaranteed to be an
independent datum.

Our model that astrophotographers tend to place their subjects in the
center of their images ignores the fact that \emph{conjunctions} of
astronomical objects on the sky are often targets of interest.  We
suspect that we see this effect in our data set.  When Comet Holmes
passed near the California Nebula (NGC 1499), many photographers
captured the conjunction.  In \figref{fig:traj} this overdensity of
images can be seen near the marked location of NGC 1499.  The nebula
appeared below the comet (at lower Dec), and it appears that our
inferred comet trajectories have been pulled down as a result.

In some sense, this project is a citizen-science project, because it
does science with data generated by non-scientists.  However, it is
very different from projects like \project{SETI@Home}
(\citealt{seti@home}) because it makes use of participants'
intelligence, not just hardware.  It is very different from projects
like \project{GalaxyZoo} (\citealt{galaxyzoo}) because it makes use of
specialized astronomy knowledge among the participants; one must be
a relatively avid astronomer to usefully contribute.  It is very
different from the projects of the \project{AAVSO} (\url{http://aavso.org})
or \project{MicroFUN} (\citealt{microfun}) because the observers
observed for reasons (for all we know) completely unrelated to our
scientific goals.  It is different from all of these projects in that
the participants contributed unwittingly.

\begin{figure}
\raggedright\centering
\thumbnail{0.24}{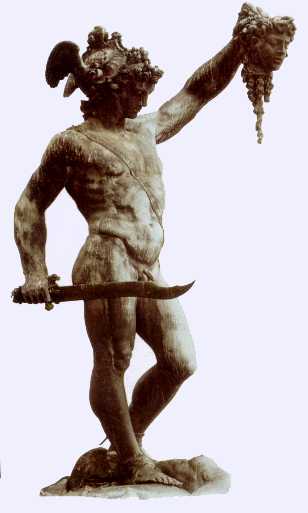}{gods}
%
\shumbnail{0.24}{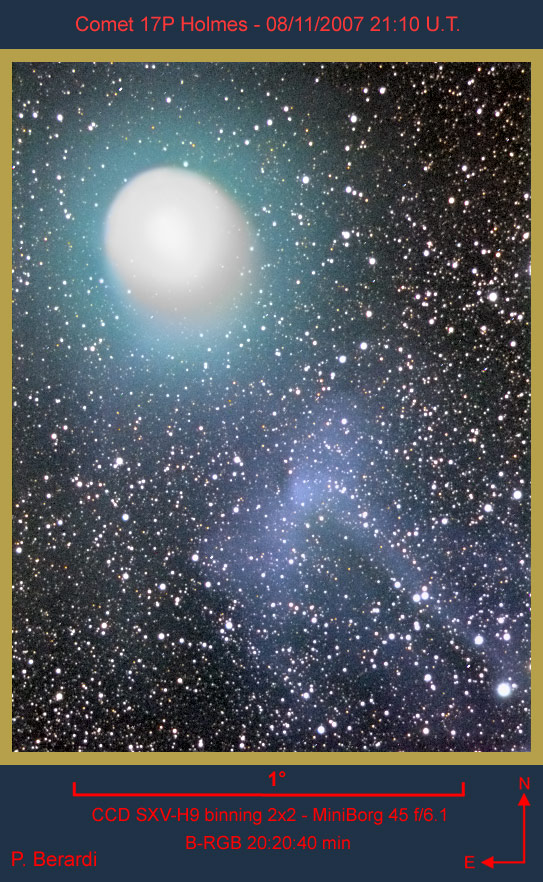}{berardi}
%
\thumbnail{0.24}{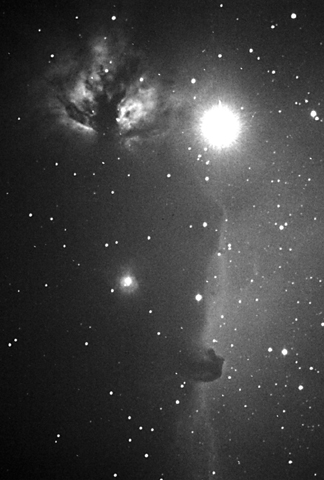}{aai}
%
\shumbnail{0.24}{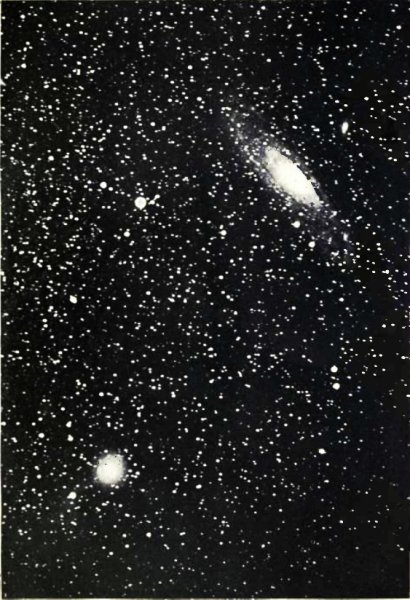}{barnard}
%
\shumbnail{0.24}{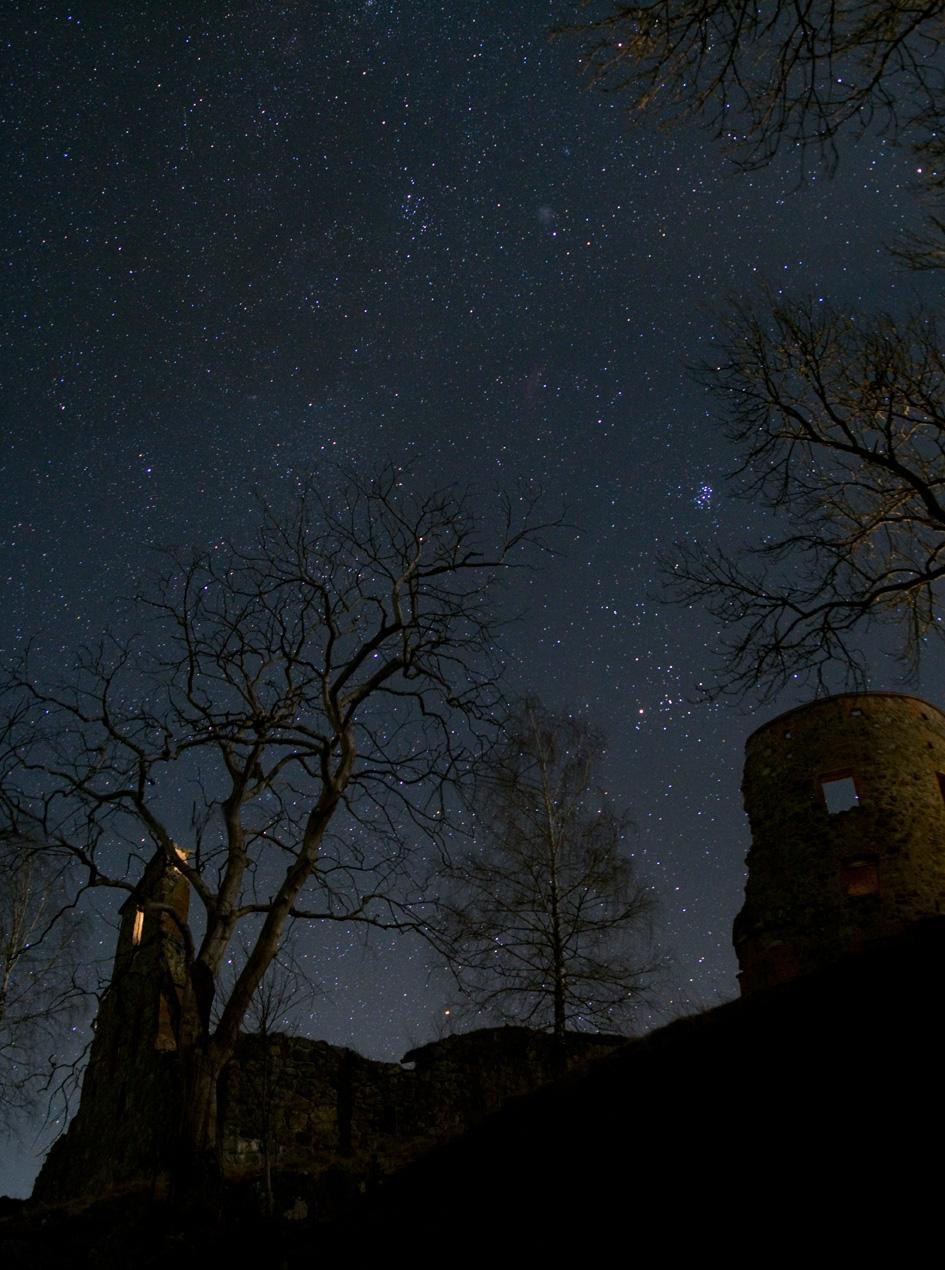}{heden}
%
\shumbnail{0.24}{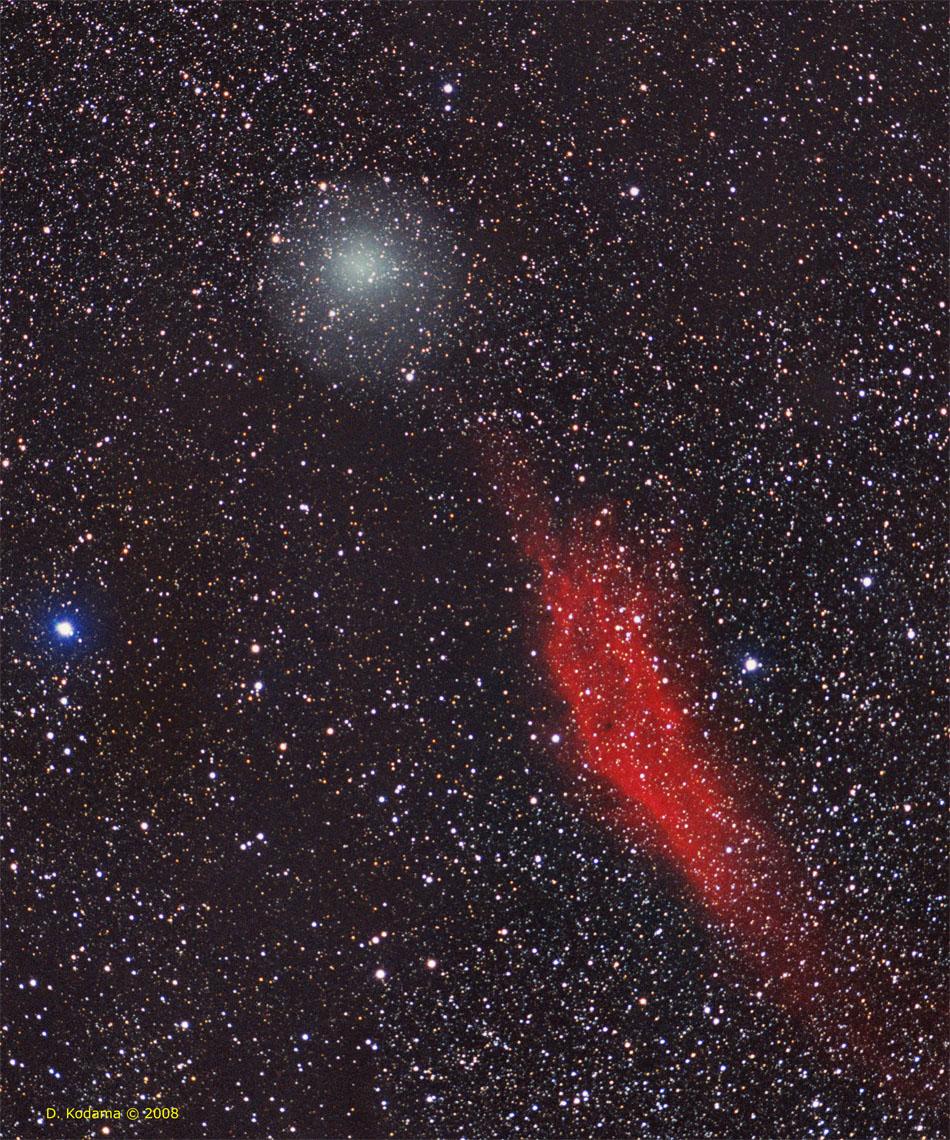}{kodama}
%
\shumbnail{0.24}{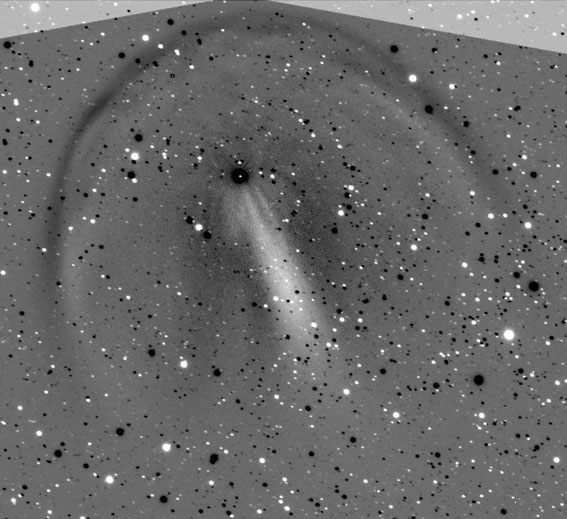}{toc}
%
\shumbnail{0.24}{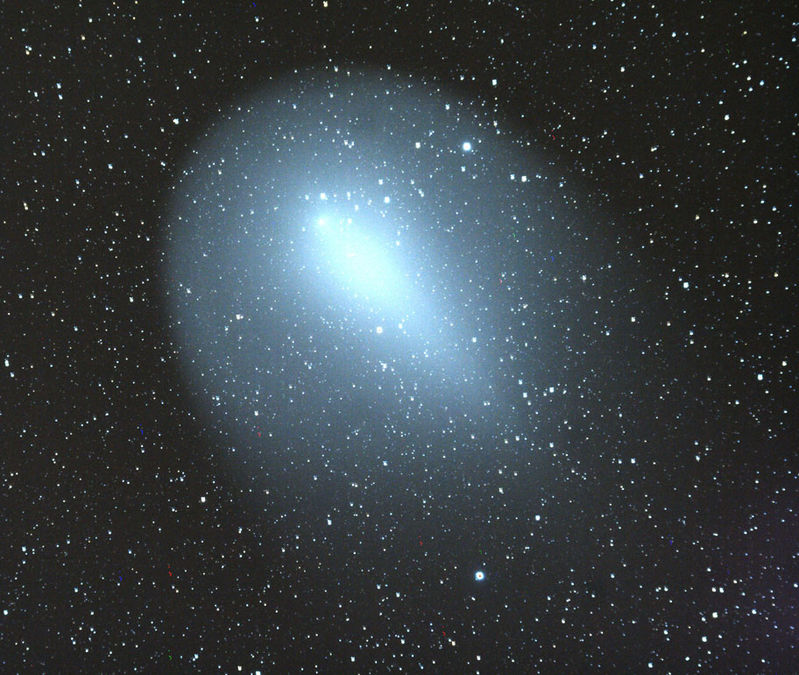}{saunders}
%
\thumbnail{0.24}{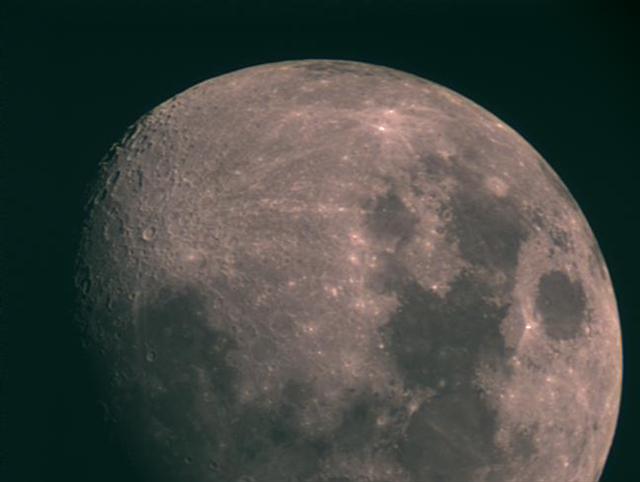}{aldrich}
%
\thumbnail{0.24}{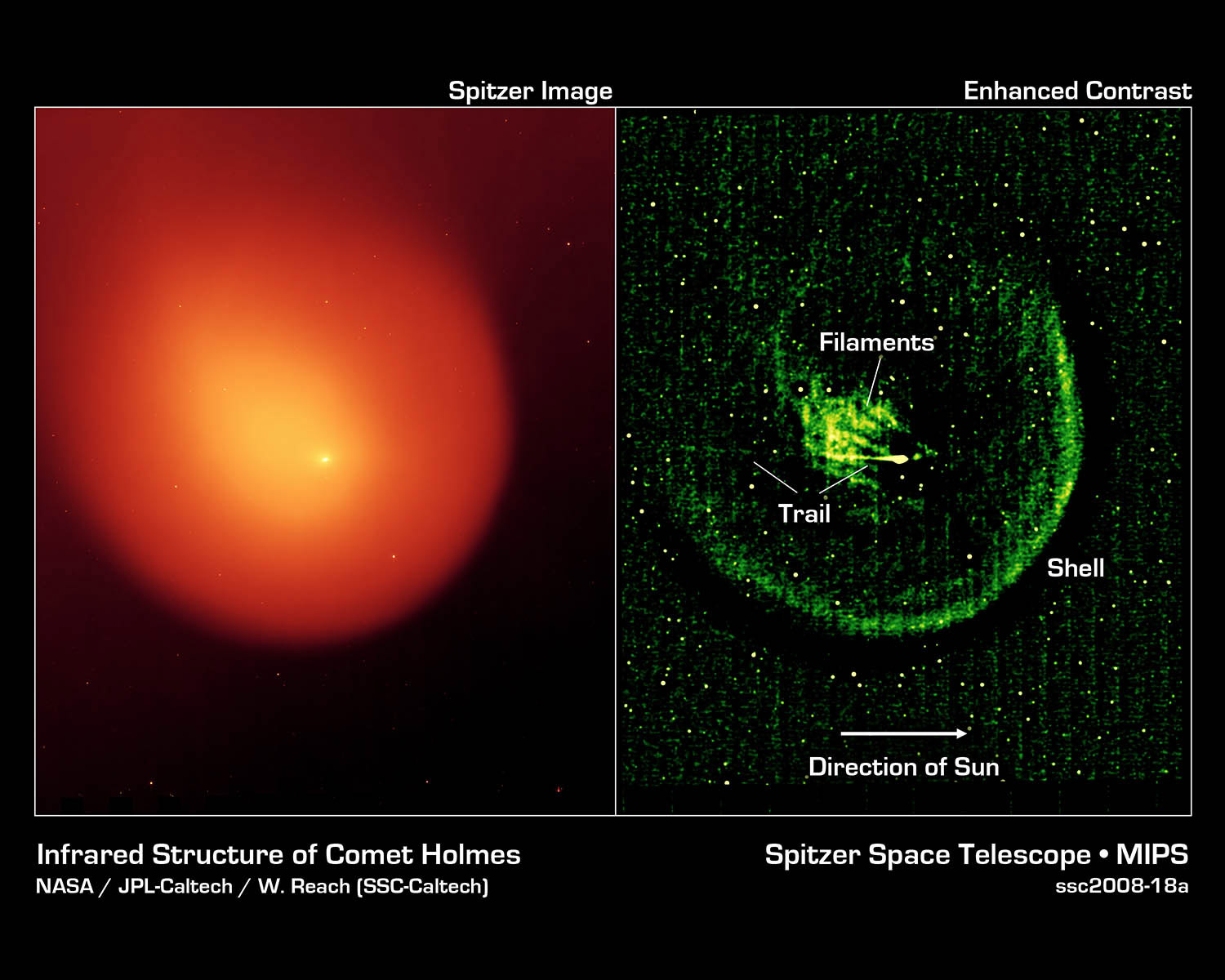}{reach}
%
\thumbnail{0.24}{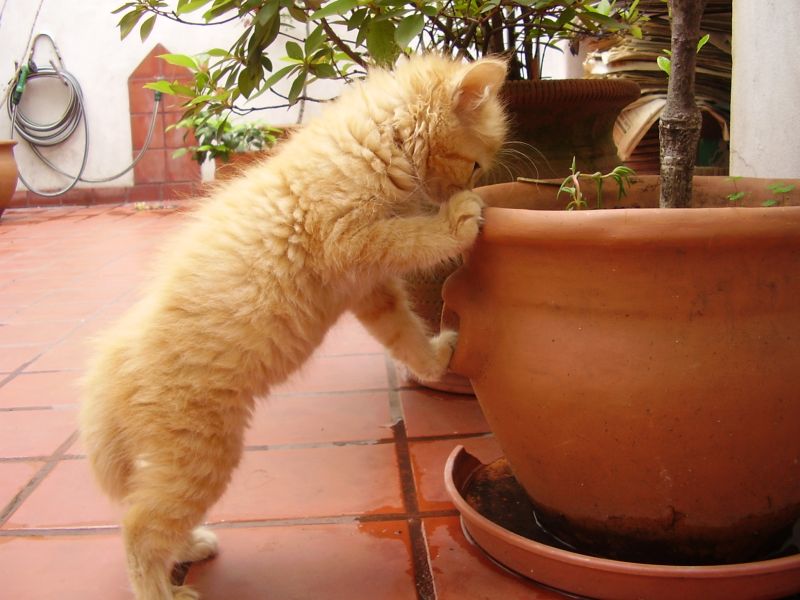}{kitty}
%
\thumbnail{0.24}{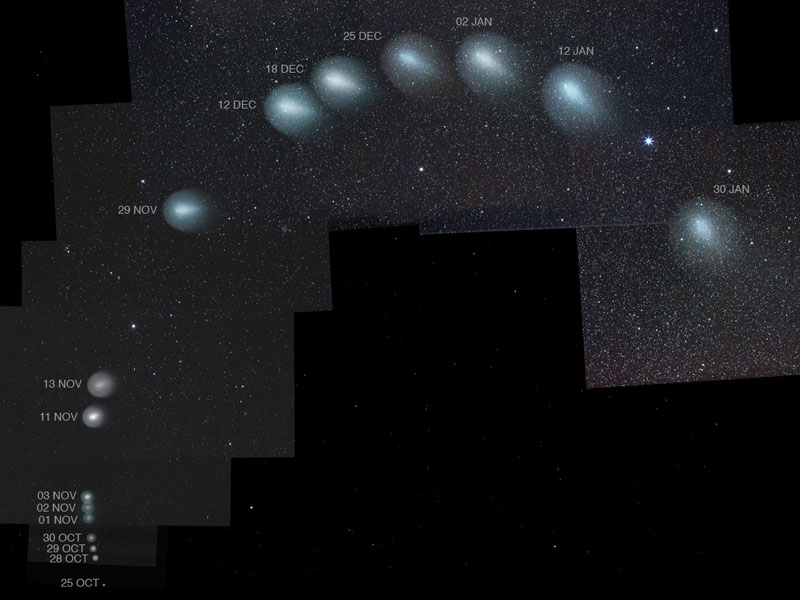}{pane}
%
\shumbnail{0.24}{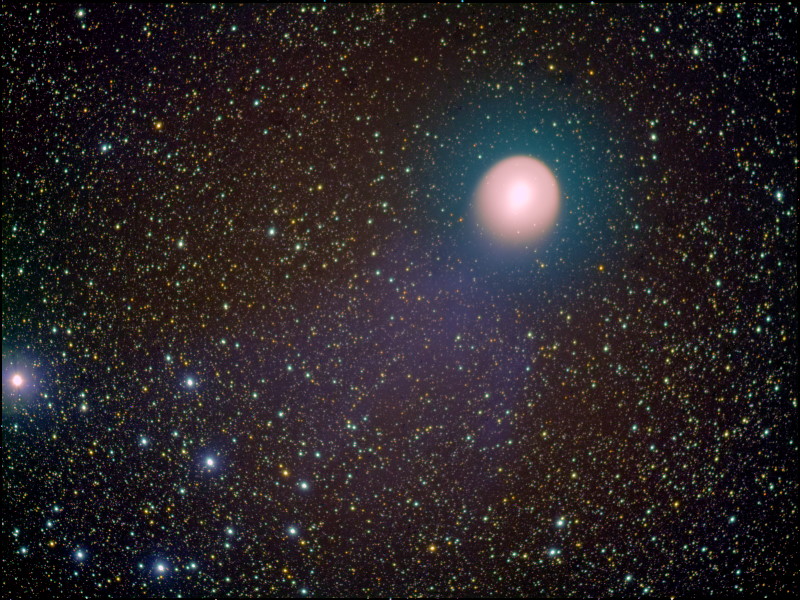}{allred}
%
\shumbnail{0.24}{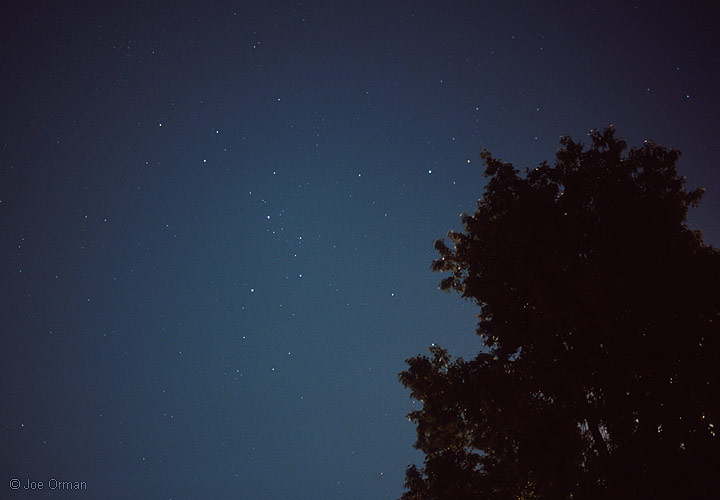}{orman}
%
\thumbnail{0.24}{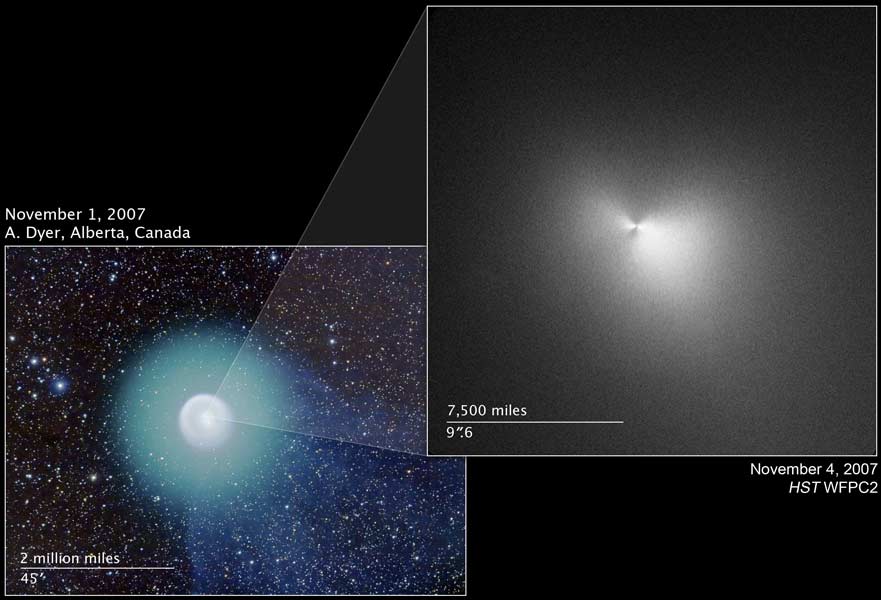}{dyer}
%
\shumbnail{0.24}{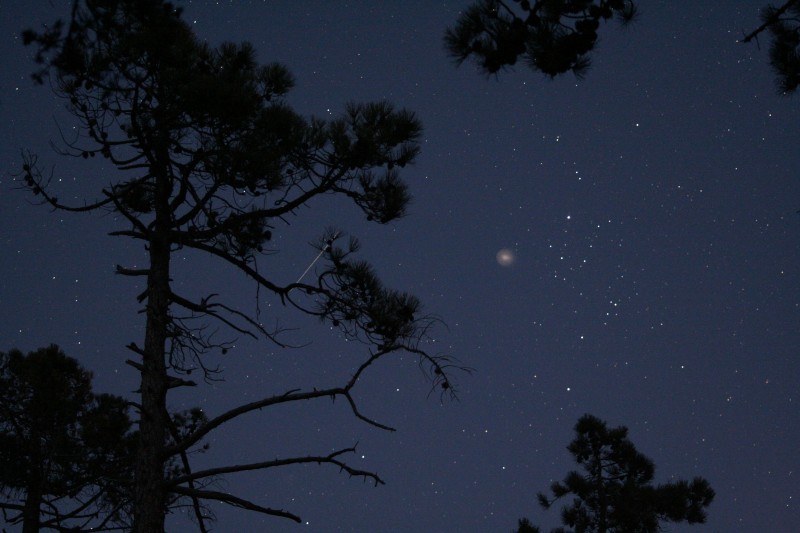}{jacques}
%
\shumbnail{0.24}{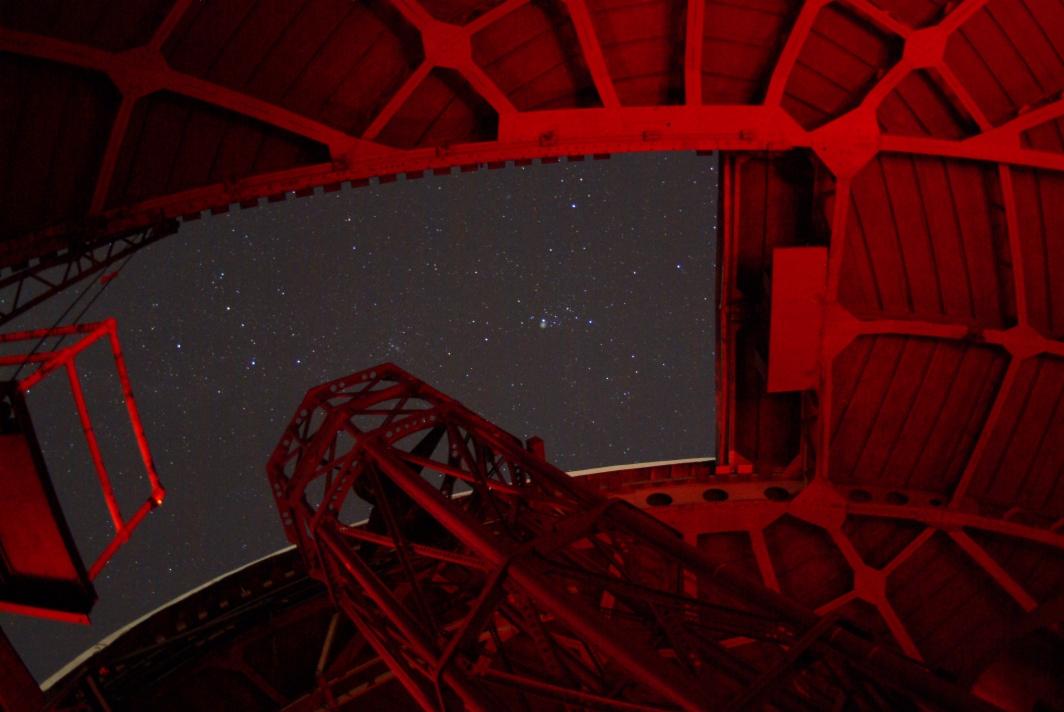}{westlake}
%
\thumbnail{0.24}{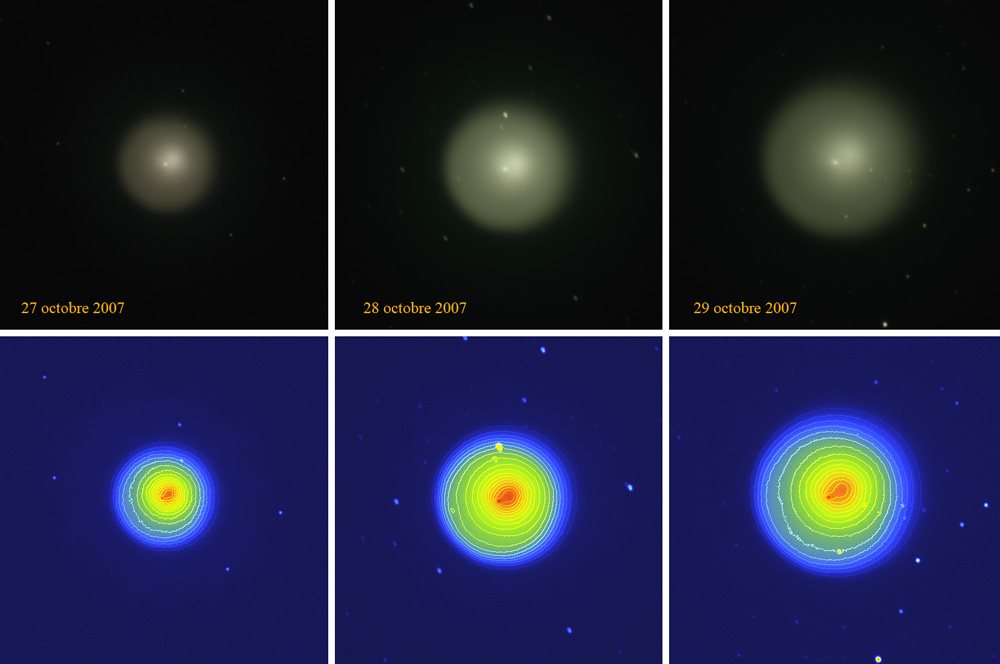}{zoll}
%
\thumbnail{0.24}{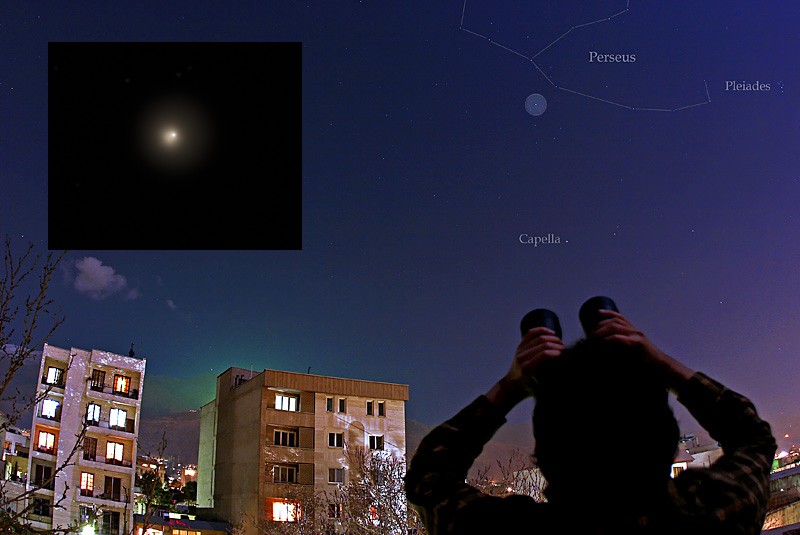}{tafreshi}
%
\shumbnail{0.24}{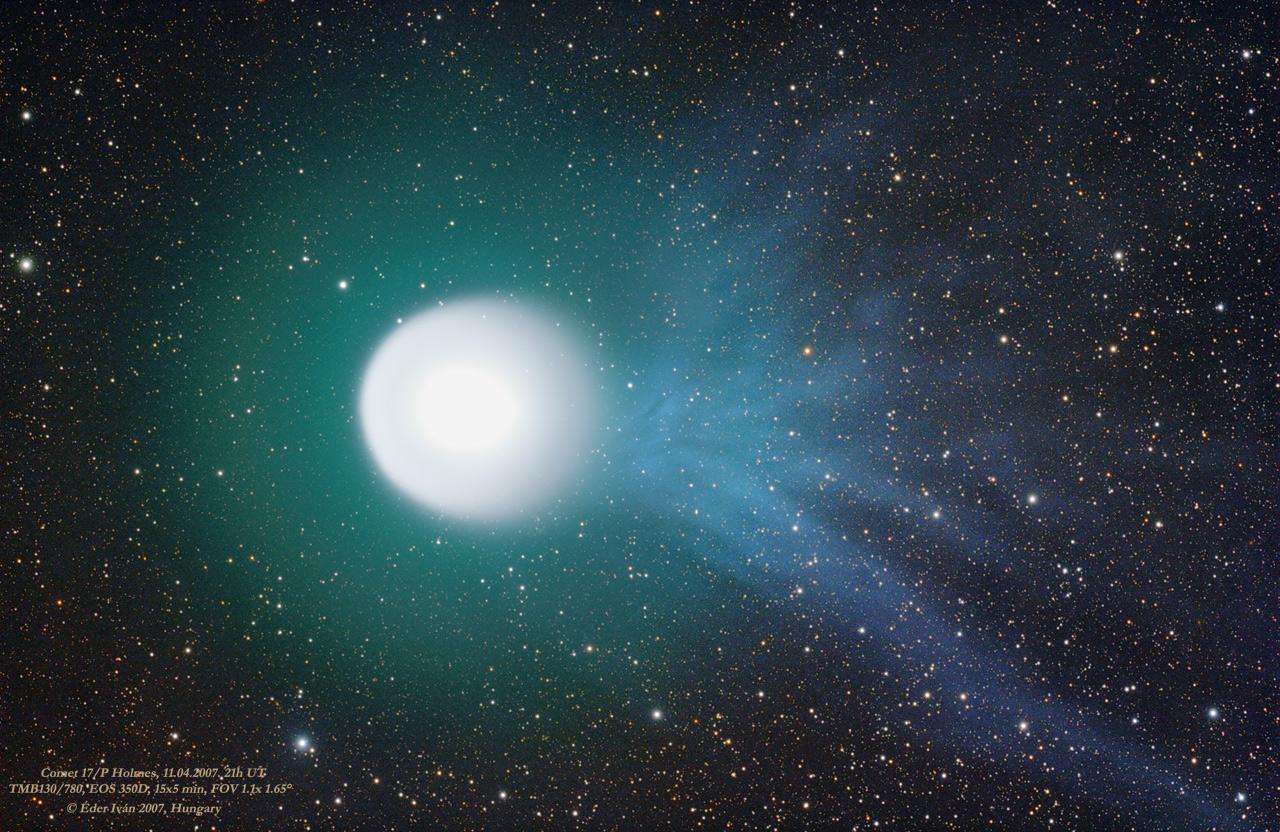}{eder}
%
\shumbnail{0.24}{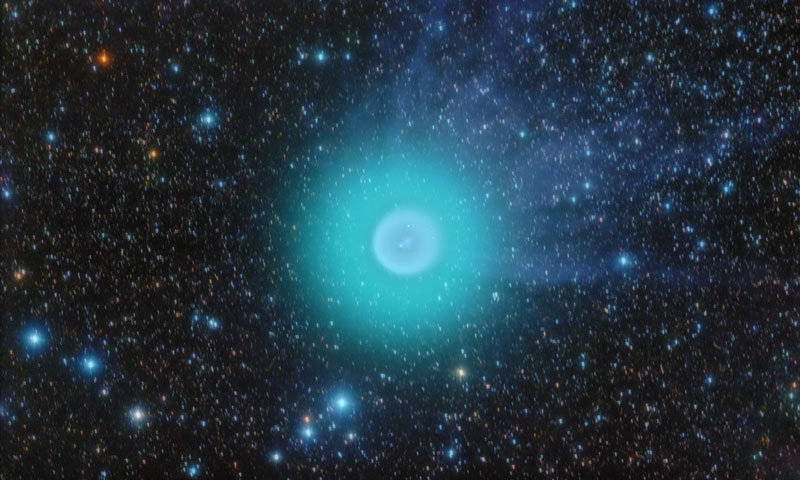}{lamadrid}
%
\thumbnail{0.32}{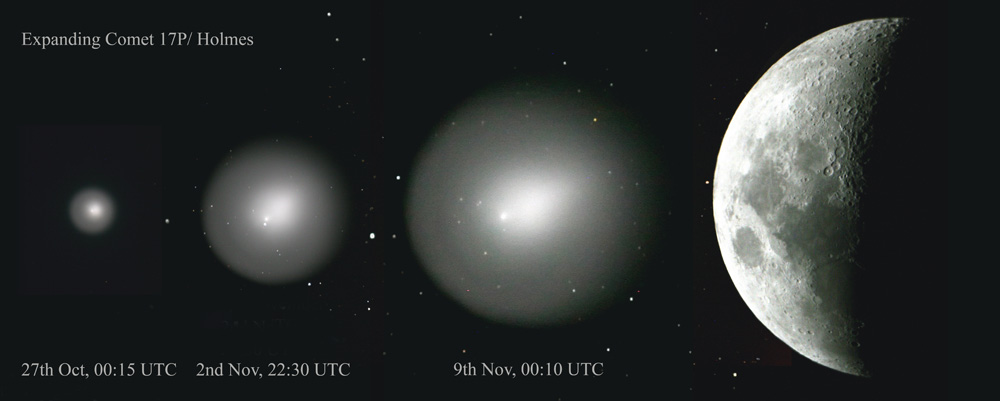}{boeckel}
%
\thumbnail{0.49}{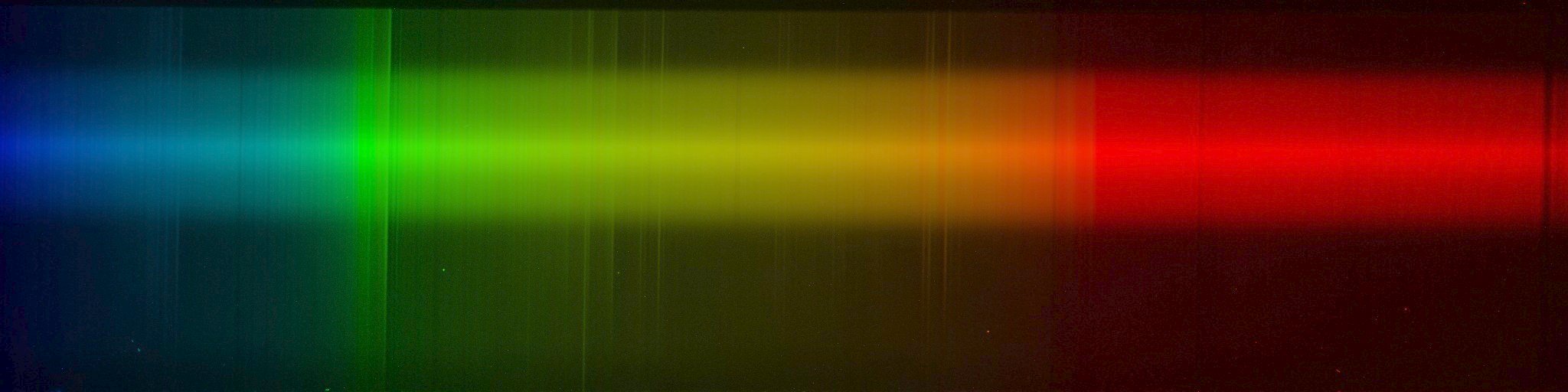}{barry}
%
\caption{Example images\ldots [caption next page]\label{fig:examples}}
\end{figure}

One interesting and ill-understood aspect of a citizen-science project
of this type---where the participants are not aware that they are
involved---relates to giving proper credit and obtaining proper
permissions to use the images.  We obtained permission to show the
images shown in \figref{fig:examples} but we did not even attempt to
get any permissions for the majority of the $2241$ images we touched
in the analysis.  One encouraging lesson from this project is that the
photographers we \emph{did} contact were very supportive: Not one
rejected our request for permissions; typical responses expressed
enthusiasm about being involved in a scientific paper; the majority
asked to see the manuscript when it appears; some sent updated images
or suggestions about which images to use; and a few offered details
about the data analysis and processing that was performed.  A less
encouraging lesson is that Web image search APIs have an uncertain
future: The \YWS\ API is being decommissioned, as is the
\project{Google Image Search} API.

The biggest lesson is that there is enormous information about
astronomy available in uncurated non-professional images on the Web.
We have only scratched this surface.  Think how much better we could
have done if we had gone into the images and actually made some
attempt at \emph{detecting} the comet!  \figref{fig:footprints} shows
that there is far more information inside the images than in just the
footprints.  \figref{fig:hyakutake} shows that there is a similarly
informative body of images of Comet C/1996 B2 (Hyakutake).  We have
also noticed that there are thousands of images of the Orion Nebula on
\flickr\ alone, and thousands more elsewhere on the Web; the joint
information in this body of images (about the nebula and about
time-domain activity therein) must be staggering.  Perhaps this is not
surprising given the large amount of telescope aperture and detector
area owned by avid photographers.  We have learned that we can do
high-quality quantitative astrophysics with images of unknown
provenance on the Web.  Is it possible to build from these images a
true sky survey?  We expect the answer is ``yes''.

\acknowledgements We are very sad that Sam Roweis (Toronto, Google,
NYU) was not here to collaborate on this project, after co-creating
with us \An; he influenced every aspect of this project.  We obtained
ideas, feedback, or code from Michael Blanton (NYU), Jo Bovy
(NYU/IAS), Daniel Foreman-Mackey (NYU), Jonathan Goodman (NYU), Fengji
Hou (NYU), Iain Murray (Edinburgh), and Christopher Stumm
(Microsoft/Etsy).  We offer sincere thanks to our anonymous reviewer
for very detailed and thoughtful comments that spurred many
improvements to the manuscript.  We benefited from the activity of a
very large number of astrophotographers and Web citizens, but in
particular we got permission to show images from John F.~Pane,
Stephane Zoll, Thorsten Boeckel, Vincent Jacques, Babek Tafreshi, Per
Magnus Hed\'en, Jimmy Westlake, Vicent Peris, Paolo Berardi, Fay
Saunders, Dave Kodama, Tyler Allred, Joe Orman, Ivan Eder, Ray Shapp,
Flemming R.~Ovesen, Torben Taustrup, and Bruce Card.  This work was
supported in part by NASA (grant NNX08AJ48G), the NSF (grant
AST-0908357), and a Research Fellowship of the Humboldt Foundation.
This research made use of the \project{SAO/NASA Astrophysics Data
  System}, the \YWS\ service, the Jet Propulsion Lab
\project{Horizons} service, the \project{Python} programming language,
and open-source software in the \project{numpy}, \project{scipy},
\project{matplotlib}, \project{emcee}, \project{pYsearch}, and
\project{WCSlib} projects.  All code and data used in this project are
available from the authors upon request.

\clearpage

\clearpage
\addtocounter{figure}{-1}
\begin{figure}
\caption[thumbnails]{[figure previous page]
  Example images from the image search.
  Here we have highlighted the diversity of the images; the
  majority are in fact high-quality, narrow-field images of the comet,
  such as \tlabel{\ref{berardi}}, \tlabel{\ref{saunders}},
  \tlabel{\ref{allred}}, and \tlabel{\ref{lamadrid}}.
  Images that did not calibrate successfully with \An\ (and therefore
  were not used as data in this study) are marked with asterisks.
  Notice that images \tlabel{\ref{heden}}, \tlabel{\ref{jacques}}, and \tlabel{\ref{westlake}}
  all were successfully calibrated and were used in the analysis.
  Image \tlabel{\ref{gods}} shows a statue of Perseus, a constellation through which Comet Holmes
  traveled during its 2007 approach.
  %
  %
  Image \tlabel{\ref{barnard}} is from the year 1892, shortly after
  the discovery of Comet Holmes.  Although it was calibrated and used
  in our analysis, this image was identified as an outlier since we
  are fitting only the 2007 apparition of the comet.
  Image \tlabel{\ref{kodama}} includes the California nebula (NGC 1499).
 %
  %
  Credits:
  \tlabel{\ref{gods}}~copyright 2000--2005 Gods, Heroes, and Myth (\url{http://www.gods-heros-myth.com});
  \tlabel{\ref{berardi}}~Paolo~Berardi;
  \tlabel{\ref{aai}}~Amateur Astronomers, Inc.~Research Committee (\url{http://asterism.org});
  \tlabel{\ref{barnard}}~Edward~Emerson~Barnard (\citealt{ball});
  \tlabel{\ref{heden}}~copyright P.-M.~Hed\'en (\url{http://www.clearskies.se});
  \tlabel{\ref{kodama}}~copyright Dave~Kodama (\url{http://astrocamera.net});
  \tlabel{\ref{toc}}~TOC Observatory (\url{http://tocobs.org});
  \tlabel{\ref{saunders}}~copyright Fay~Saunders;
  \tlabel{\ref{aldrich}}~Bruce Card, Aldrich Astronomical Society, Worcester MA;
  \tlabel{\ref{reach}}~NASA, JPL-Caltech, W.~Reach (SSC-Caltech);
  \tlabel{\ref{kitty}}~copyright Juli\'an~Cantarelli;
  \tlabel{\ref{pane}}~copyright 2007, 2008 John~F.~Pane (\url{http://holmes.johnpane.com});
  \tlabel{\ref{allred}}~copyright Tyler~Allred (\url{http://allred-astro.com});
  \tlabel{\ref{orman}}~Joe~Orman;
  \tlabel{\ref{dyer}}~NASA, ESA, and H.~Weaver (The Johns Hopkins University Applied Physics Laboratory), and A.~Dyer, Alberta, Canada;
  \tlabel{\ref{jacques}}~copyright Vincent Jacques (\url{http://vjac.free.fr/skyshows});
  \tlabel{\ref{westlake}}~Jimmy~Westlake, Colorado Mountain College;
  \tlabel{\ref{zoll}}~Stephane~Zoll (\url{http://astrosurf.com/zoll});
  \tlabel{\ref{tafreshi}}~Babak~Tafreshi / TWAN (\url{http://twanight.org});
  \tlabel{\ref{eder}}~Ivan~Eder (\url{http://eder.csillagaszat.hu});
  \tlabel{\ref{lamadrid}}~Vicent~Peris (OAUV), Jos\'e~Luis~Lamadrid (CEFCA);
  \tlabel{\ref{boeckel}}~copyright Thorsten~Boeckel (\url{http://tboeckel.de});
  \tlabel{\ref{barry}}~D.~J.~Barry, Department of Astronomy, Cornell University.
  }
\end{figure}

\clearpage
\begin{figure}
\begin{center}
\includegraphics[width=0.45\textwidth]{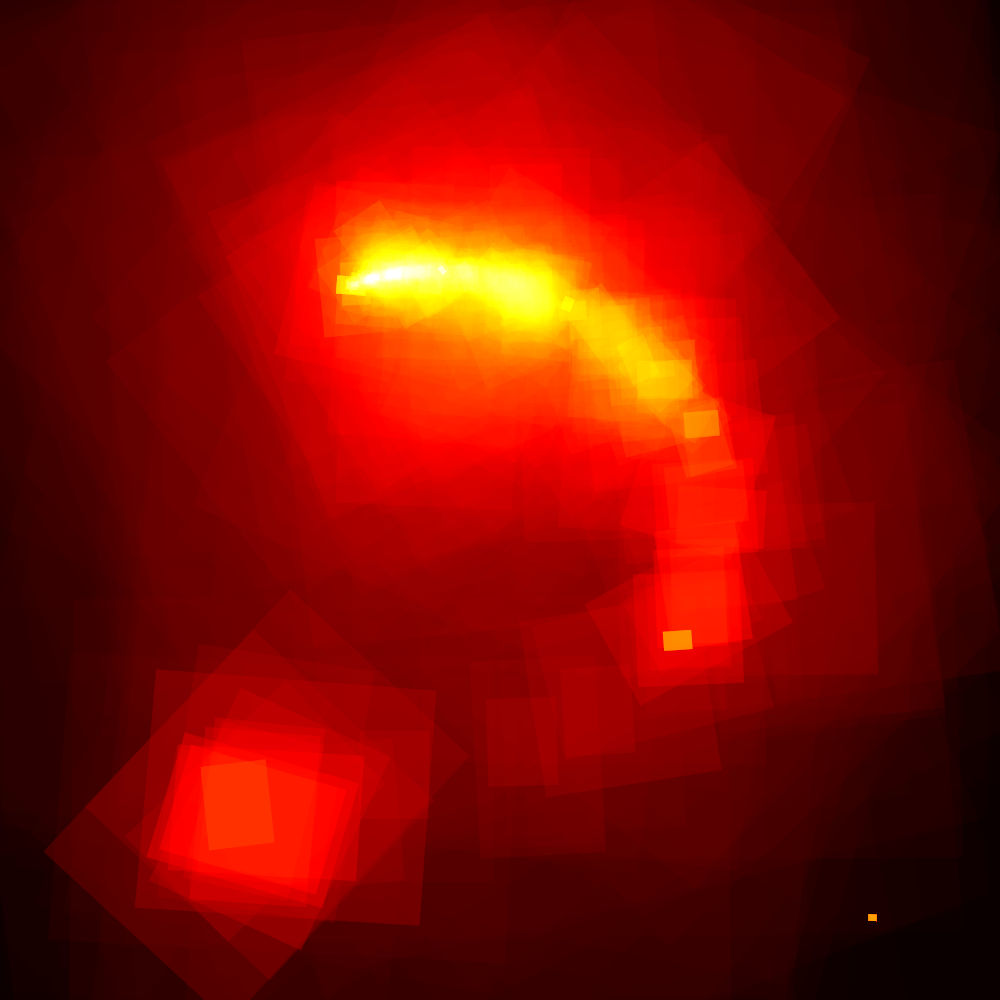}\hspace{1pt}%
\includegraphics[width=0.45\textwidth]{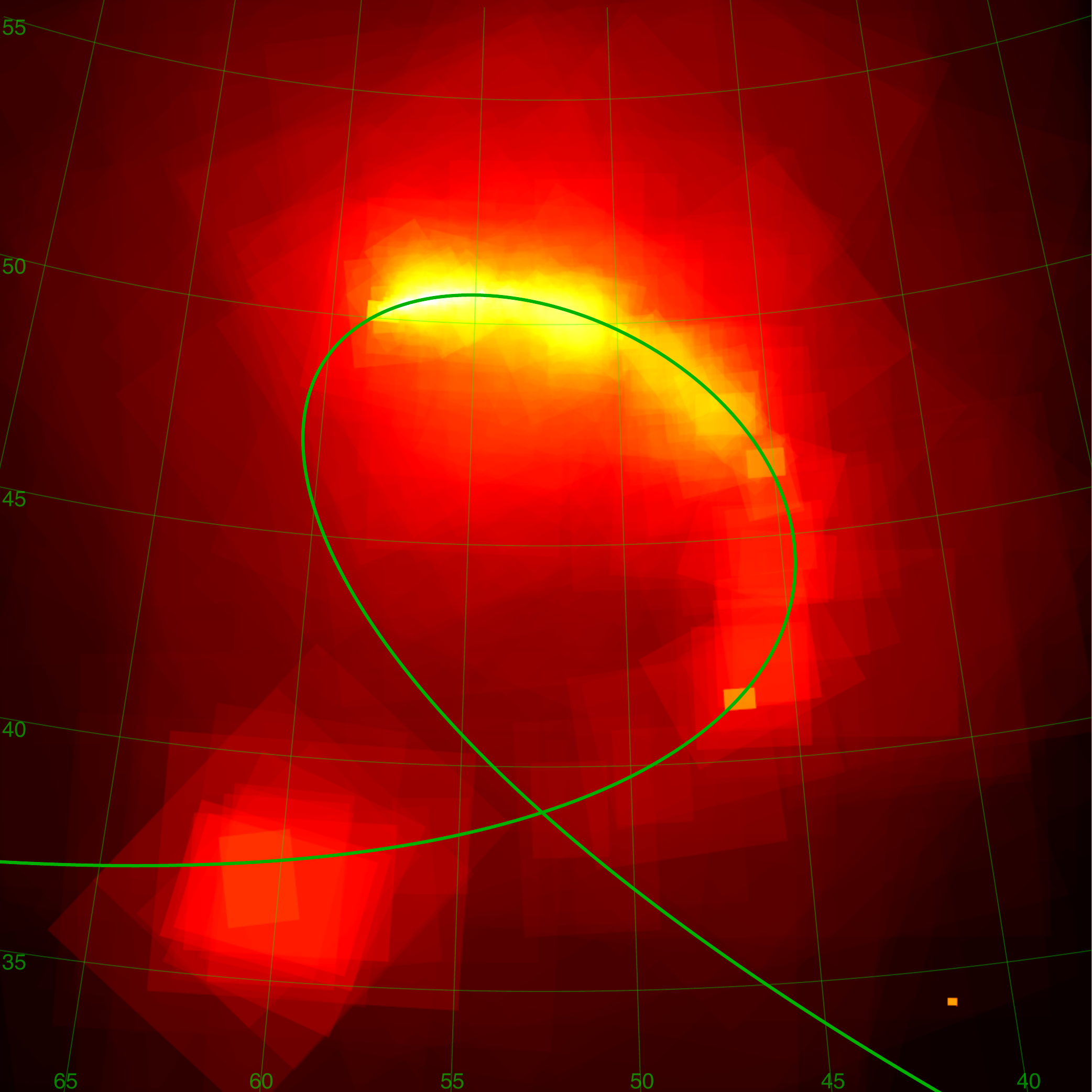}\hspace{1pt}%
\includegraphics[height=0.45\textwidth]{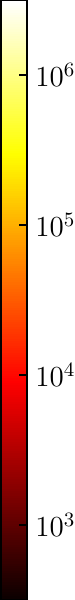}
\includegraphics[width=0.45\textwidth]{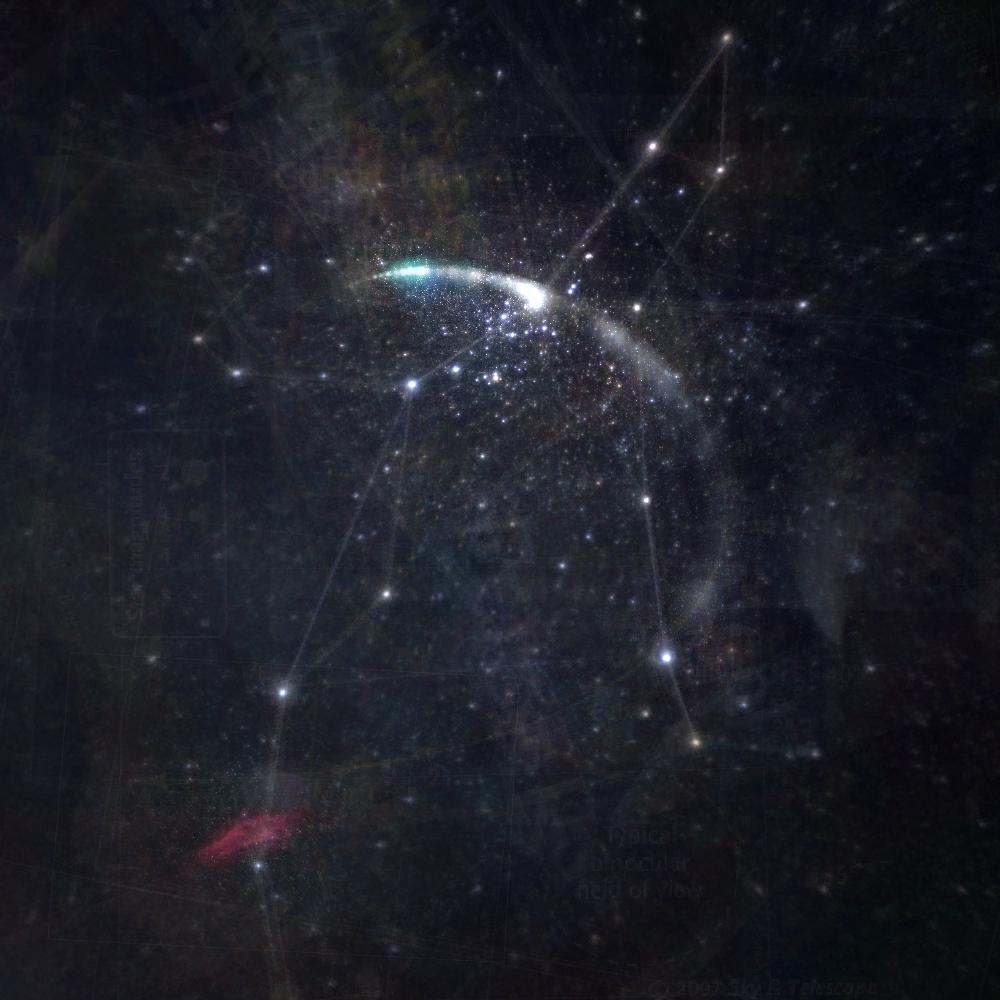}\hspace{1pt}%
\includegraphics[width=0.45\textwidth]{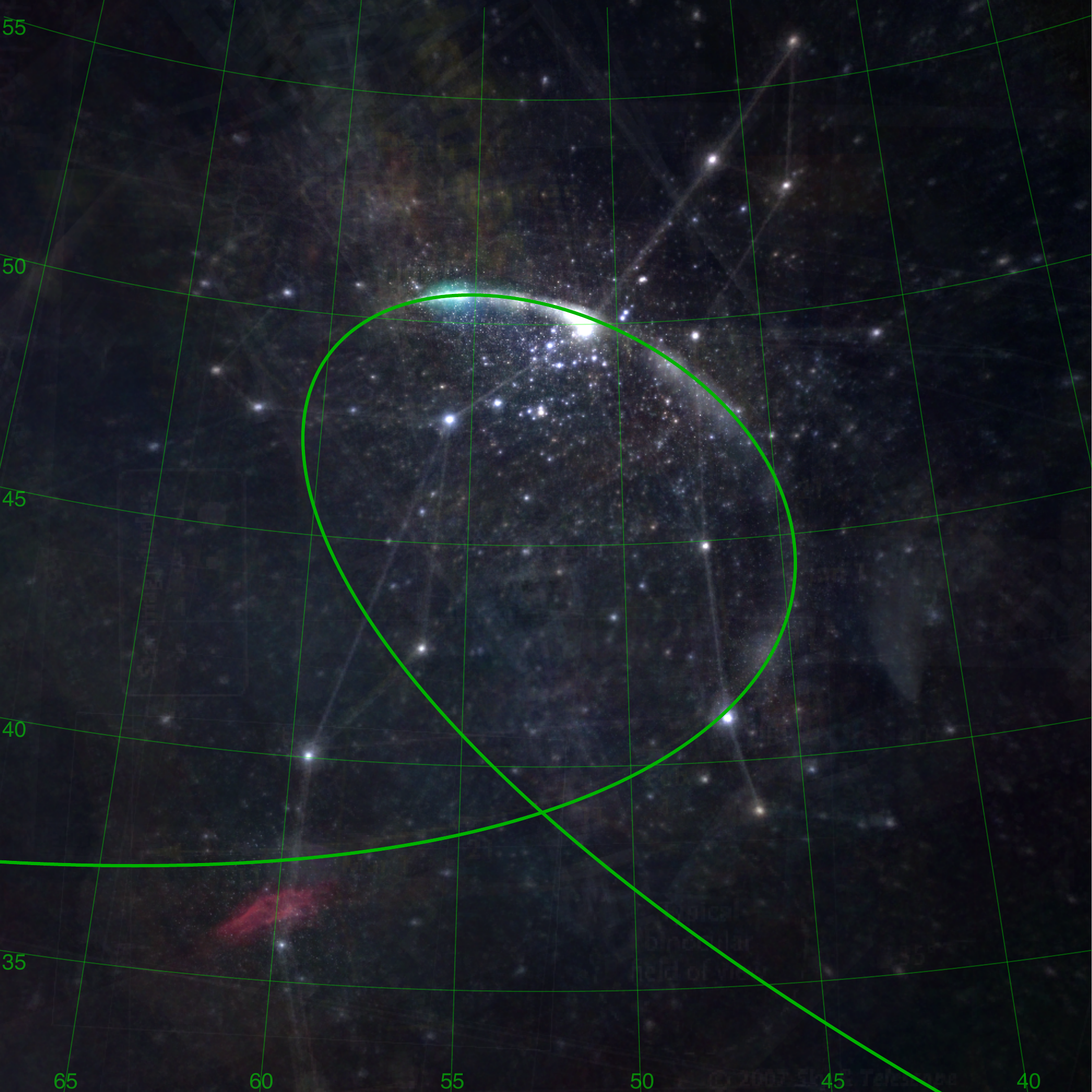}\hspace{1pt}%
\rule{0.0575\textwidth}{0pt}
\end{center}
\caption[footprints]{Images that were successfully calibrated by \An,
  aligned in celestial coordinates.  We show only the region within
  about 20 degrees of the comet's 2007 trajectory; there are 134
  images outside the plot area.  \textsl{Top:} Total pixel density of
  the images, with a log stretch.  The most heavily imaged point on
  the sky is covered by 493 images and has a pixel density of over $4$
  million pixels per square degree.  The colorbar units are pixels per
  square degree.  The right panel shows the same image as the left
  panel but with a coordinate grid and the trajectory of the JPL
  ephemeris (\citealt{jpl}) for Comet 17P/Holmes.  \textsl{Bottom:}
  Co-added images.  The co-added images show the fixed stars because
  the images have been aligned in celestial coordinates; they show
  some faint lines joining the stars because some of the images used
  in this study are diagrams of the constellations rather than plain
  photographs.  The California Nebula (NGC 1499) is visible in the
  bottom-left of the image because many photographers imaged the
  conjunction of the comet and nebula (see, for example,
  \figref{fig:examples}\tlabel{\ref{kodama}}).\label{fig:footprints}}
\end{figure}

\clearpage
\begin{figure}
\begin{center}
\includegraphics[width=0.49\textwidth]{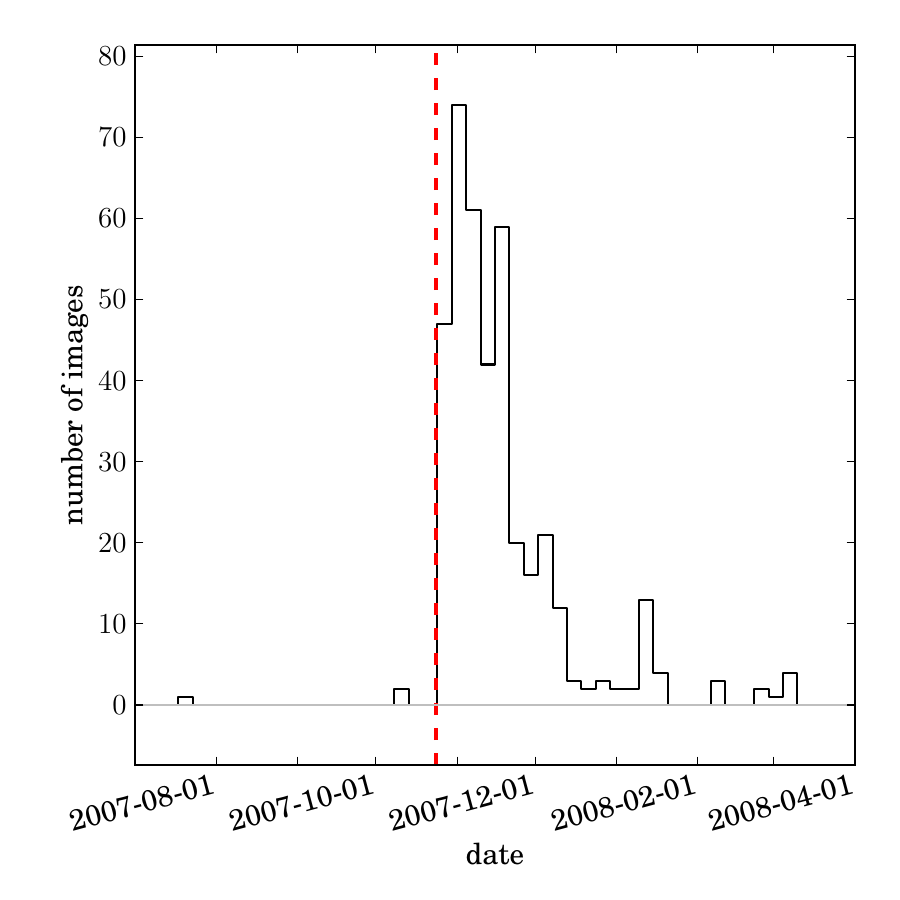}
\includegraphics[width=0.49\textwidth]{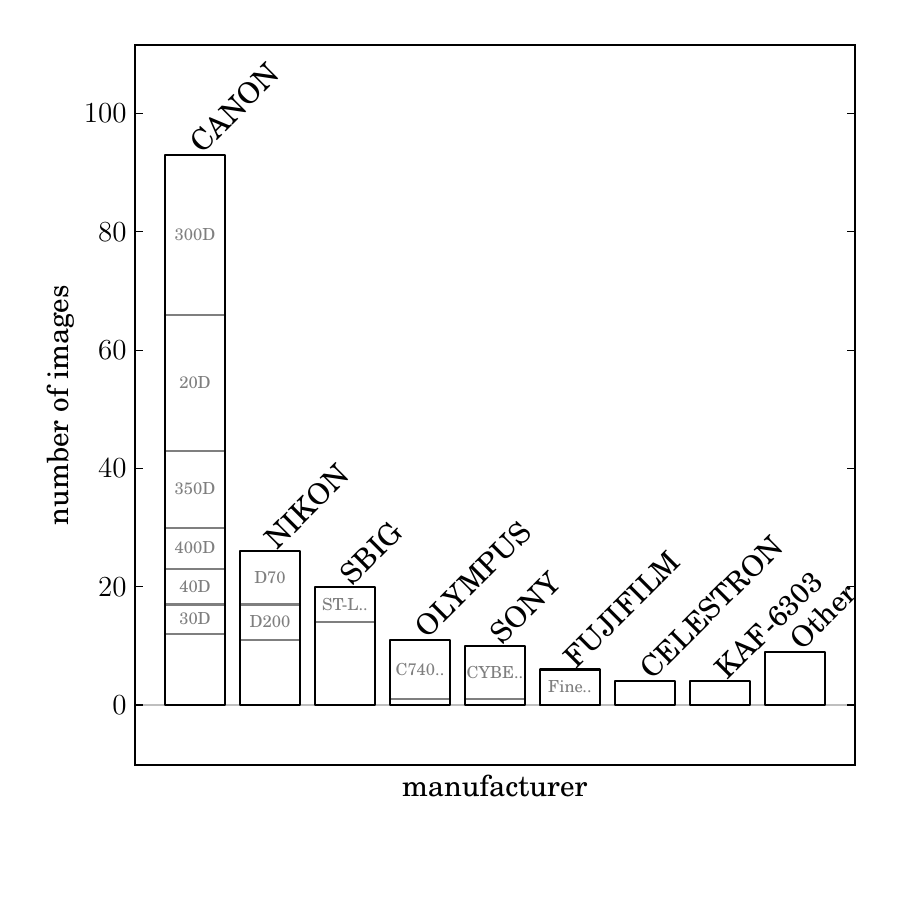}
\includegraphics[width=0.49\textwidth]{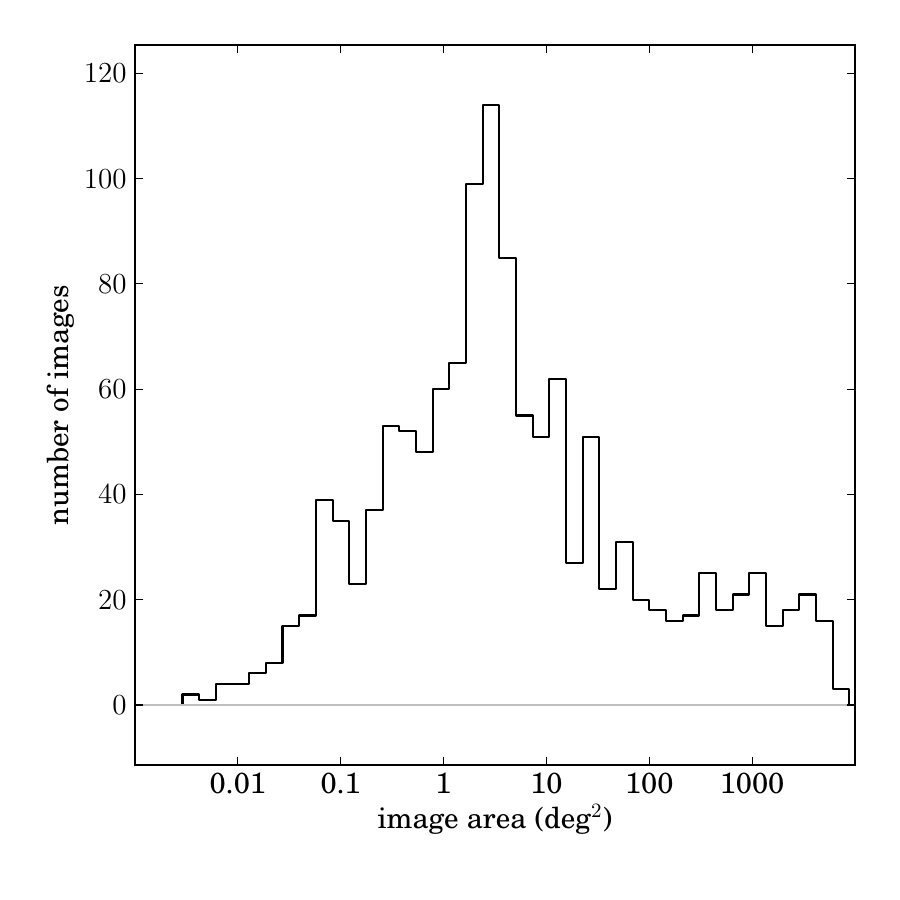}
\includegraphics[width=0.49\textwidth]{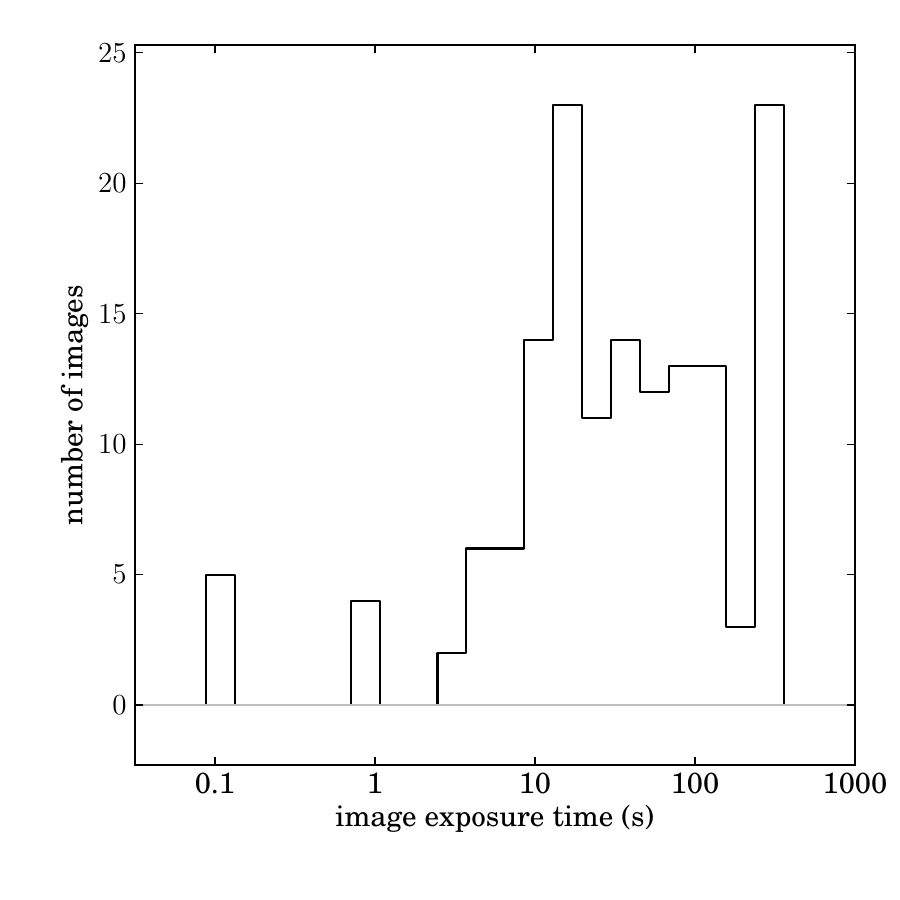}
\end{center}
\caption[meta-data]{\textsl{Top-left:} The timestamps in the EXIF
  headers of the images in our data set.  $422$ of the $1299$ images
  have timestamps; $394$ are within the range plotted here.
  The dashed line marks 2007~Oct~24, the date of the
  comet's outburst and dramatic brightening (\citealt{outburst}).
  \textsl{Top-right:} The distribution of camera manufacturers
  listed in the EXIF headers (for the 183 images containing manufacturer
  information).
  \textsl{Bottom-left:}  The distribution of image angular sizes in our
  data set, according to \An.
  \textsl{Bottom-right:} The distribution of exposure times in our
  data set, according to EXIF headers (for the 149 images containing
  an exposure time entry).\label{fig:imgstats}}
\end{figure}


\clearpage
\begin{figure}
\begin{center}
\includegraphics[width=0.6\textwidth]{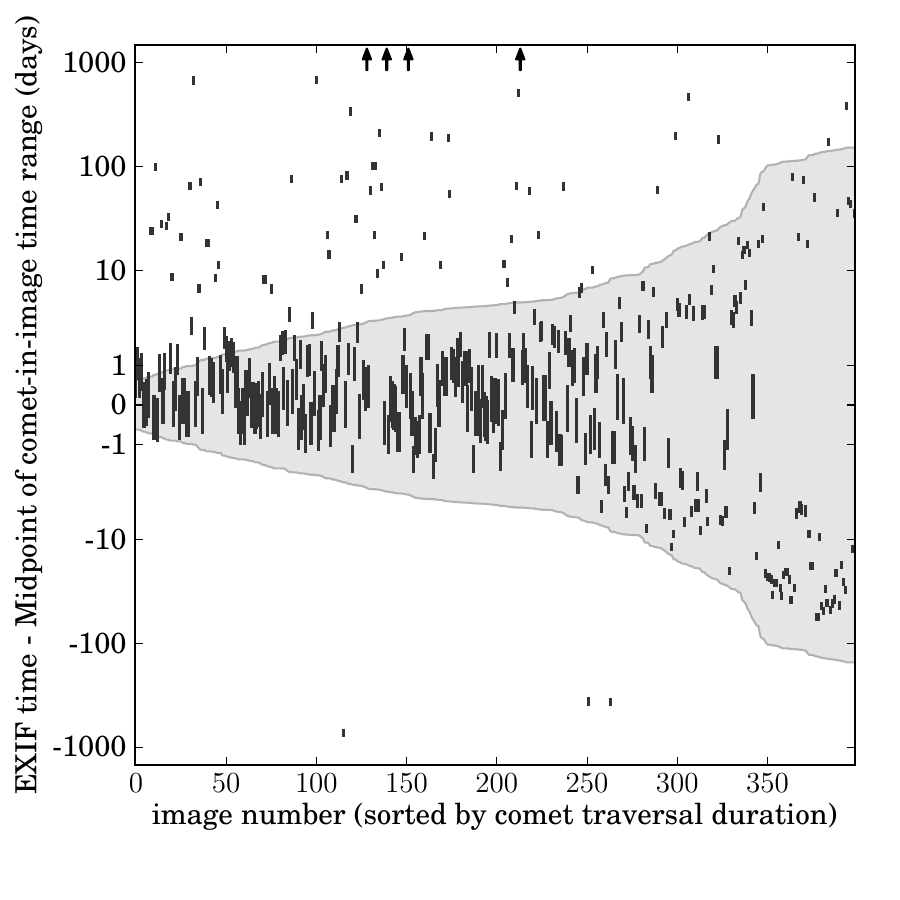}
\end{center}
\caption[exif times mostly correct]{Evaluation of the accuracy of the
  timestamps in the image EXIF headers.  For each image, we computed
  the range of times that the comet appeared inside the
  celestial-coordinate bounds of the image; that is the shaded gray
  region, which is vertically centered at zero.  The images are sorted
  so that this envelope is monotonic.  The EXIF timestamp for each
  image is shown as a bar of height one day (since EXIF has no time
  zone specification, this is our intrinsic uncertainty), plus a
  minimum size to make all the markers visible.  If the EXIF
  timestamps were correct and set to UT, the position of the bar
  within the gray region would indicate the position of the comet
  within the image.  Most of the bars touch the gray region,
  indicating that the majority of the EXIF timestamps are consistent
  (that is, the comet would indeed appear within the image at the
  stamped time), and the inconsistent times are almost all
  \emph{later}; perhaps these timestamps mark times at which the image
  was edited.  The apparent bifurcation toward the right side of the
  plot is due to the nonlinear (arcsinh) mapping we have applied to
  the vertical axis.  Of the $422$ images with timestamps, $23$ did
  not intersect the comet's trajectory (during the 2007 apparition);
  this plot shows the $399$ that did intersect the trajectory.  The
  arrows at the top of the plot show the locations of $4$ images
  outside the plot bounds.\label{fig:exiftimes}}
\end{figure}

\clearpage
\begin{figure}
\begin{center}
\includegraphics[width=0.6\textwidth]{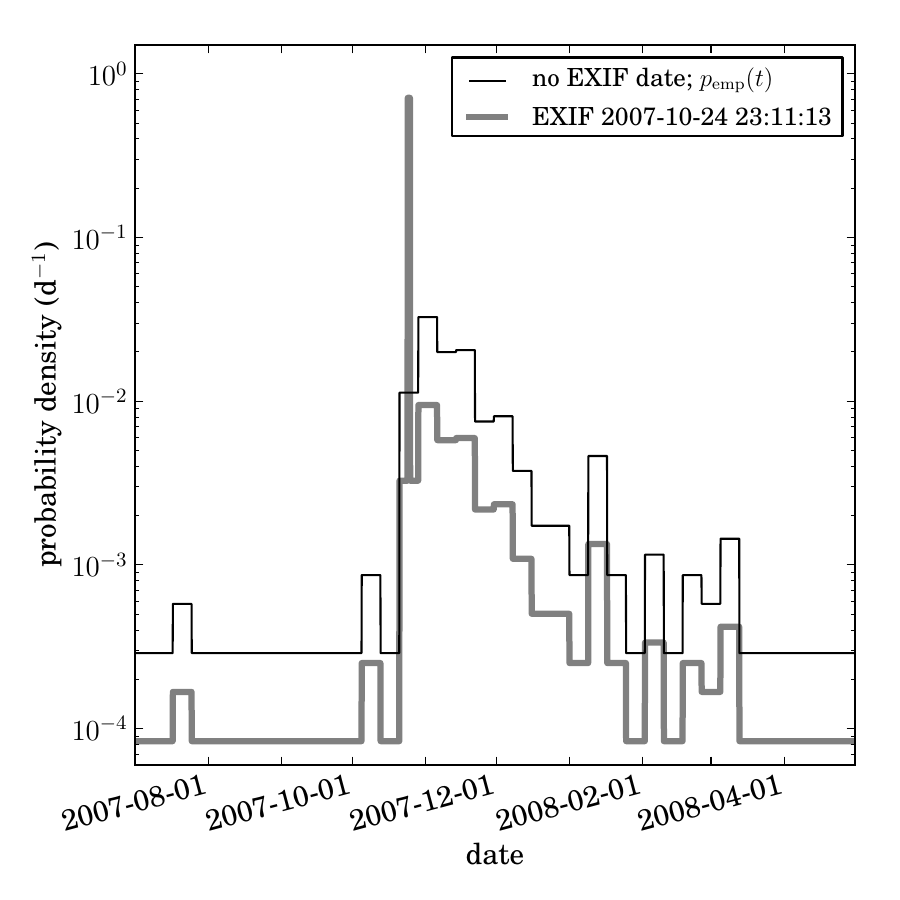}
\end{center}
\caption[exif times]{The time prior PDF $\pempirical(t)$ used for
  images with no EXIF date information (dark line) and the time prior
  PDF for an image with a particular EXIF date (lighter line).  The
  latter is a mixture of $\pempirical(t)$ and a top-hat of width one
  day centered on the EXIF date and fractional weight $\pexif = 0.71$.
  We use a width of one day because the EXIF standard does not permit
  the time zone to be specified.  The function $\pempirical(t)$ is
  based on the distribution of reported EXIF dates shown in
  \figref{fig:imgstats}; details in the
  text.\label{fig:empirical}}
\end{figure}

\clearpage
\begin{figure}
\begin{center}
\includegraphics[width=\textwidth]{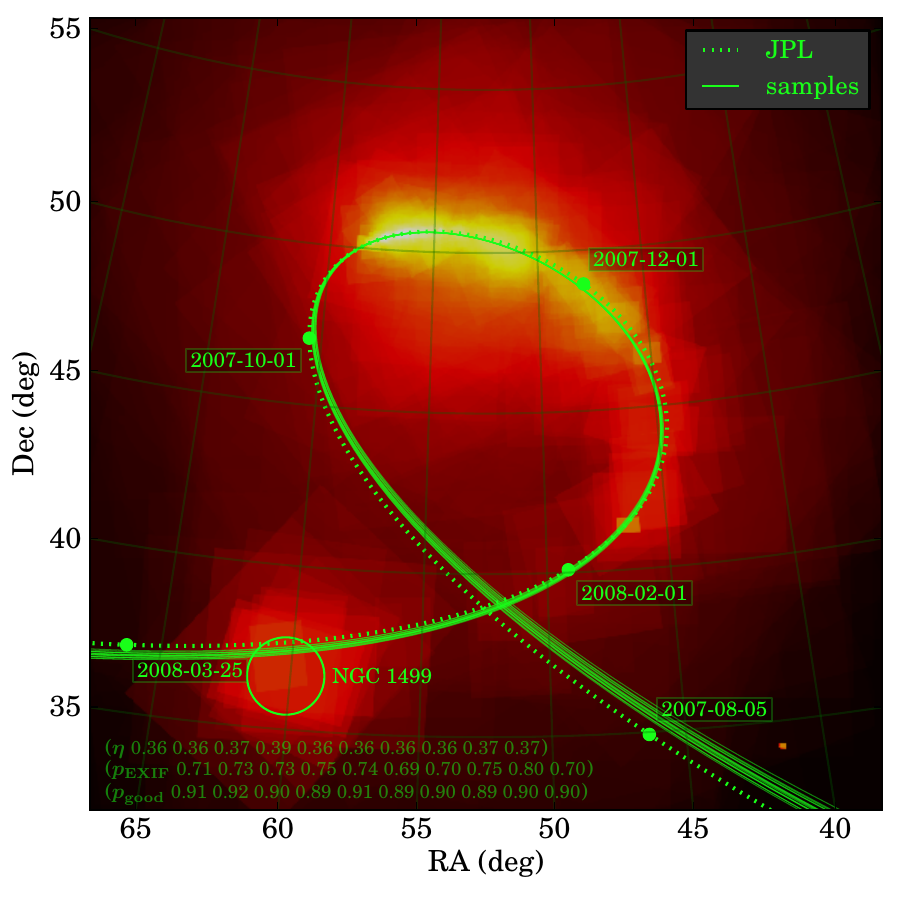}
\end{center}
\caption[results]{Image footprints with JPL ephemeris and inferred
  orbits superimposed.  The solid lines show 16 samples from the
  posterior PDF for the parameters.  The values for the
  hyperparameters $\pgood$, $\pexif$, and $\eta$ (see text for
  definitions) for some of the samples are listed at the bottom of
  the plot.  The background shows the pixel density as in
  \figref{fig:footprints}.\label{fig:traj}}
\end{figure}

\clearpage
\begin{figure}
\begin{center}
\begin{tabular}{@{}c@{}c@{}c@{}}
\includegraphics[width=0.33\textwidth]{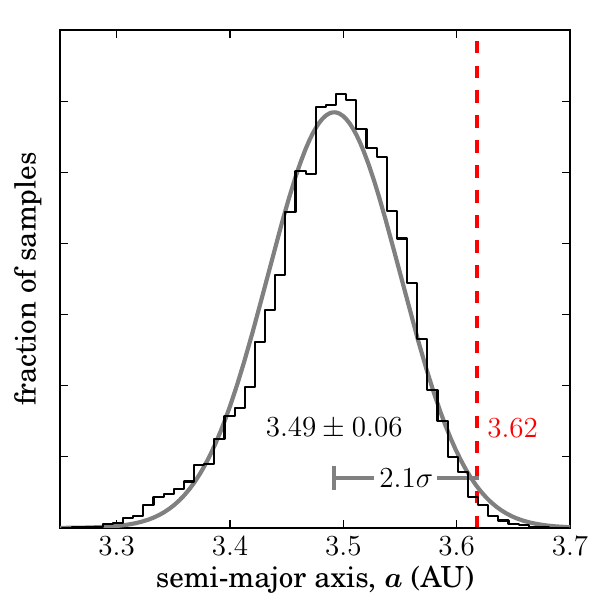} &
\includegraphics[width=0.33\textwidth]{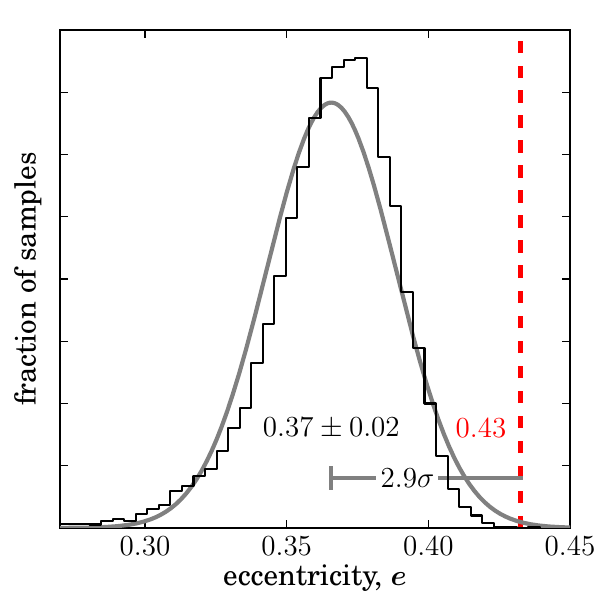} &
\includegraphics[width=0.33\textwidth]{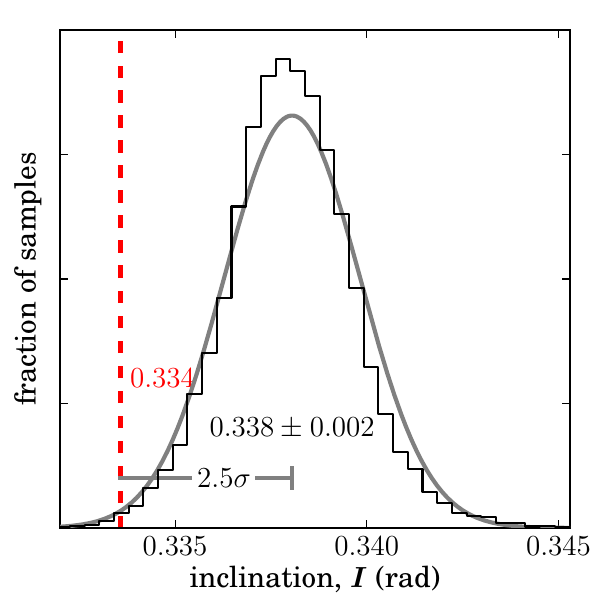} \\
\includegraphics[width=0.33\textwidth]{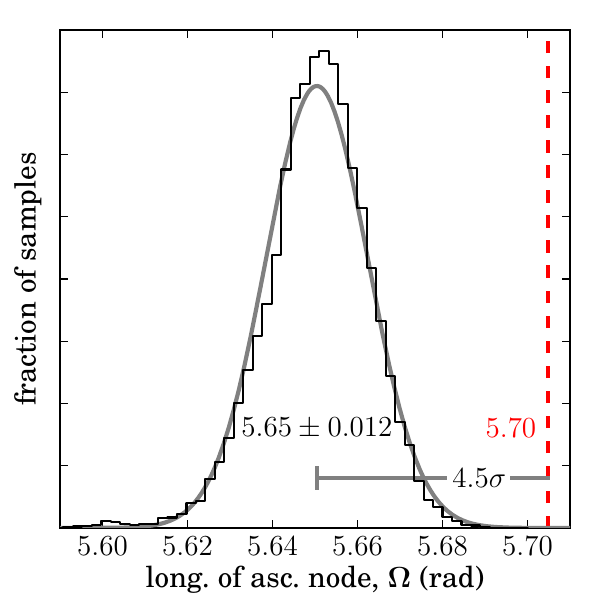} &
\includegraphics[width=0.33\textwidth]{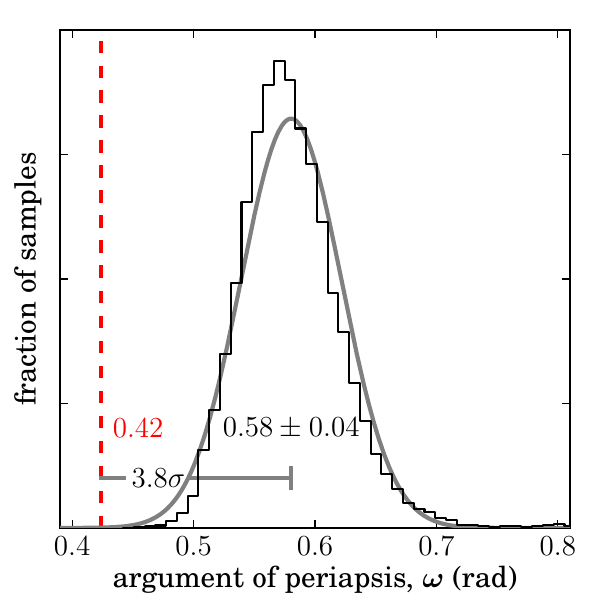} &
\includegraphics[width=0.33\textwidth]{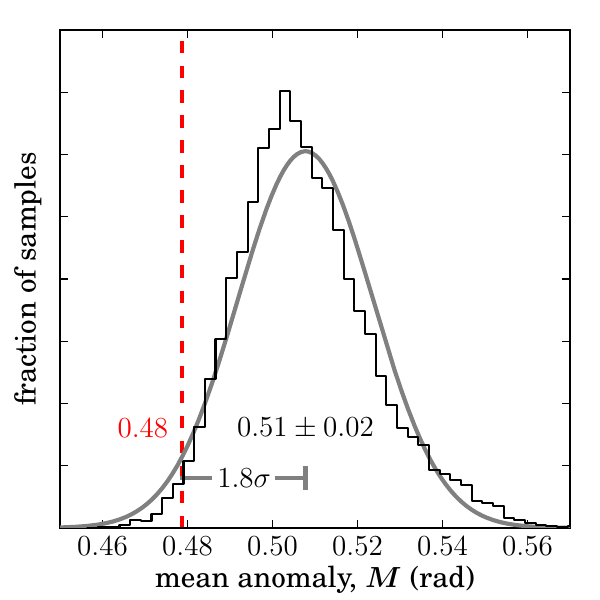} \\
\includegraphics[width=0.33\textwidth]{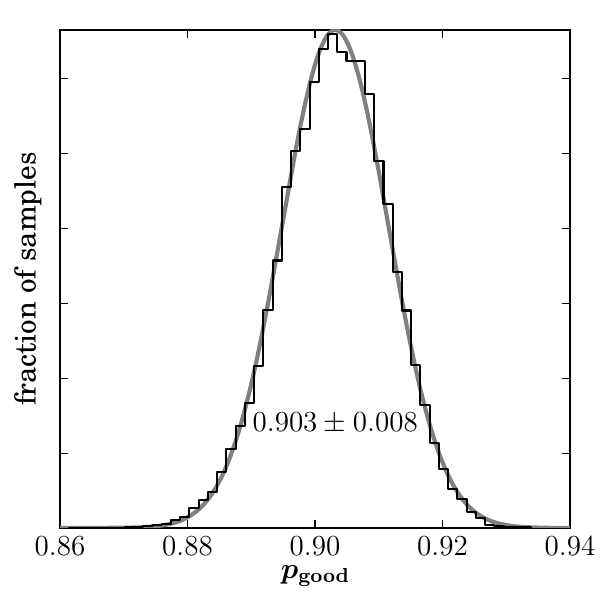} &
\includegraphics[width=0.33\textwidth]{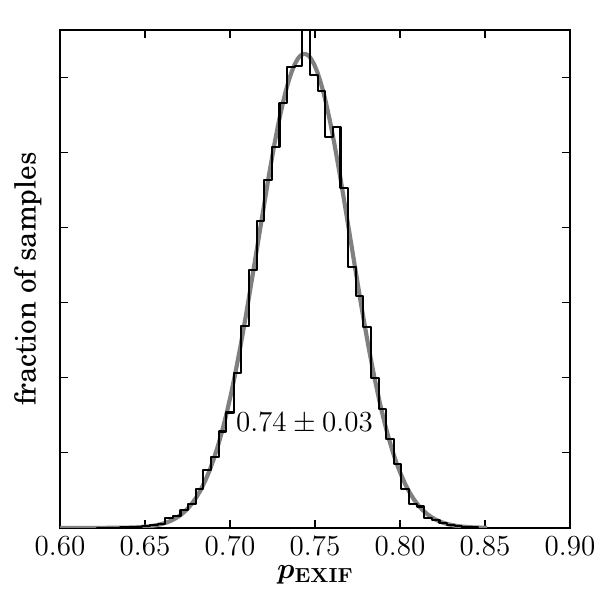} &
\includegraphics[width=0.33\textwidth]{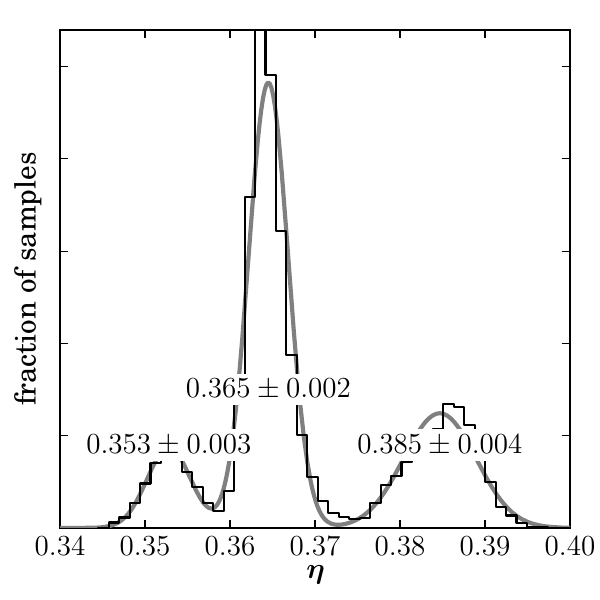}
\end{tabular}
\end{center}
\caption{\textsl{Top two rows:} Orbital parameters inferred by our
  method (histogram and best-fit Gaussian), and JPL ephemeris values
  (vertical bar).  Most of our inferred parameter distributions are a
  few standard deviations from the JPL values.  In the next figure we
  show that our inferred trajectory is closer to the JPL trajectory
  than the differences in orbital elements might suggest.
  \textsl{Bottom row:} Our inferred hyper-parameters: $\pgood$, the
  probability that an image is an image of the comet (that is, was
  generated by the foreground probability distribution $\pfg$);
  $\pexif$, the probability that a timestamp in an image EXIF header
  is correct; and $\eta$, the central fraction of the image area in
  which the comet appears.
  \label{fig:param-hists}}
\end{figure}

\clearpage
\begin{figure}
\begin{center}
\begin{tabular}{@{}c@{}c@{}}
\includegraphics[width=0.45\textwidth]{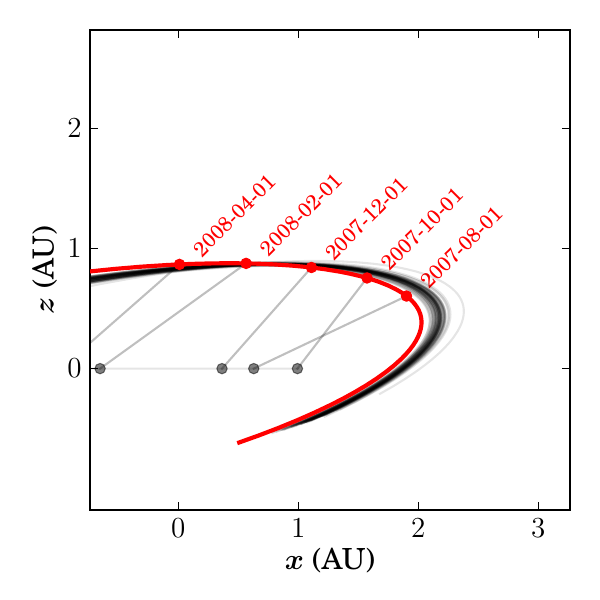} &
\\
\includegraphics[width=0.45\textwidth]{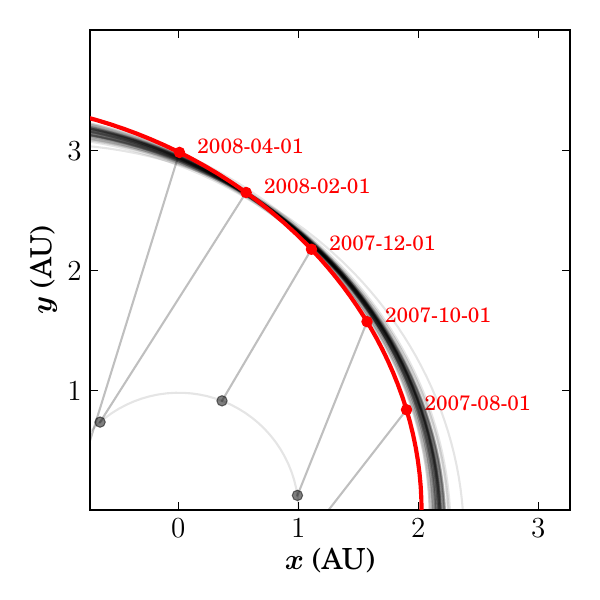} &
\includegraphics[width=0.45\textwidth]{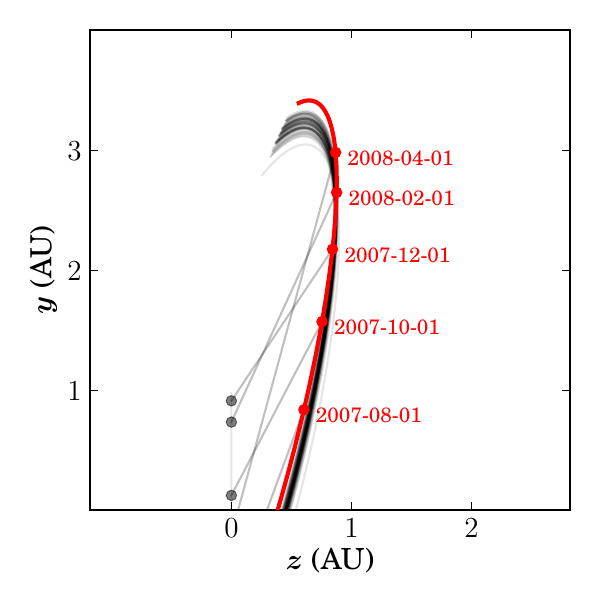}
\end{tabular}
\end{center}
\caption{Three-dimensional inferred orbit (thin lines) and JPL orbit
  (dashed line) in heliocentric equatorial coordinates.  We have plotted
  32 samples from our inferred orbit distribution to show the scatter.
  Also shown is the JPL orbit of the Earth-Moon barycenter.  Our
  inferred orbit captures the general shape of the trajectory of the
  comet, and is relatively unconstrained at early and late times (when
  we have no data). \label{fig:three-d}}
\end{figure}

\clearpage
\begin{figure}
\begin{center}
\includegraphics[width=0.49\textwidth]{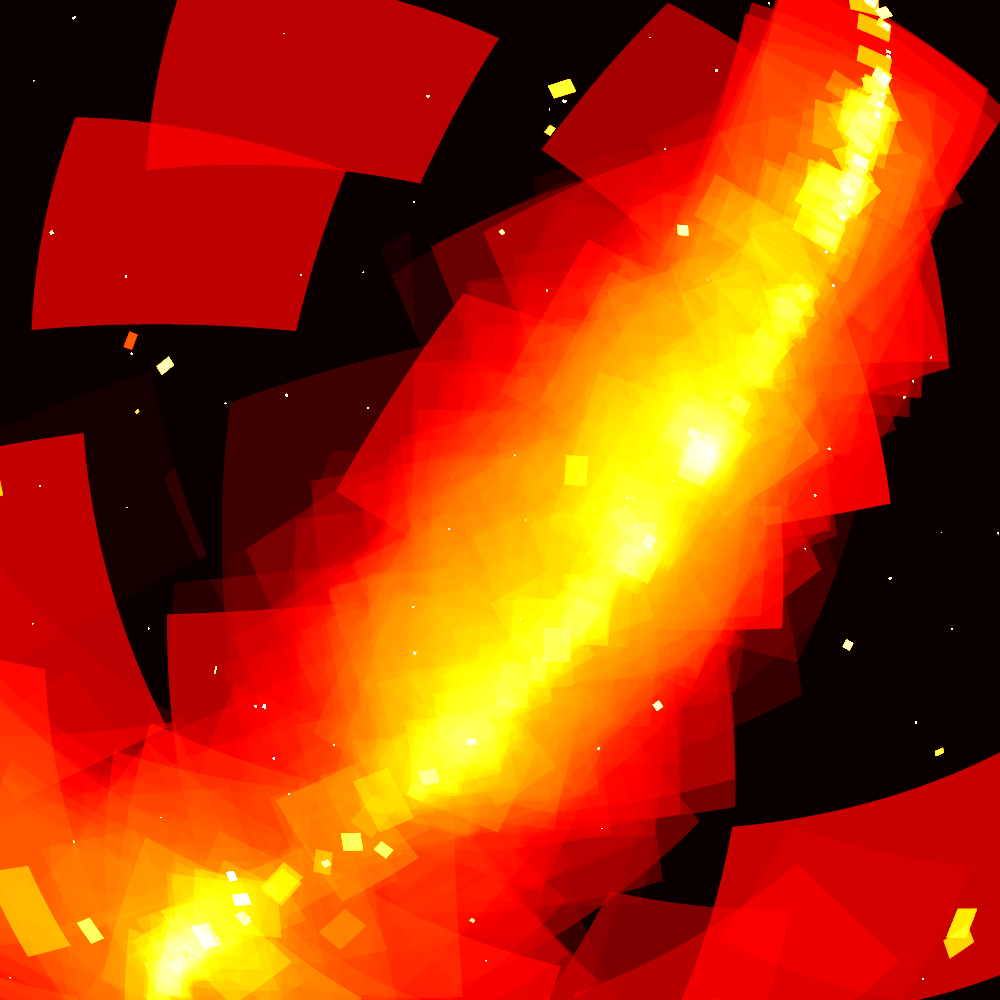}\hspace{1pt}%
\includegraphics[width=0.49\textwidth]{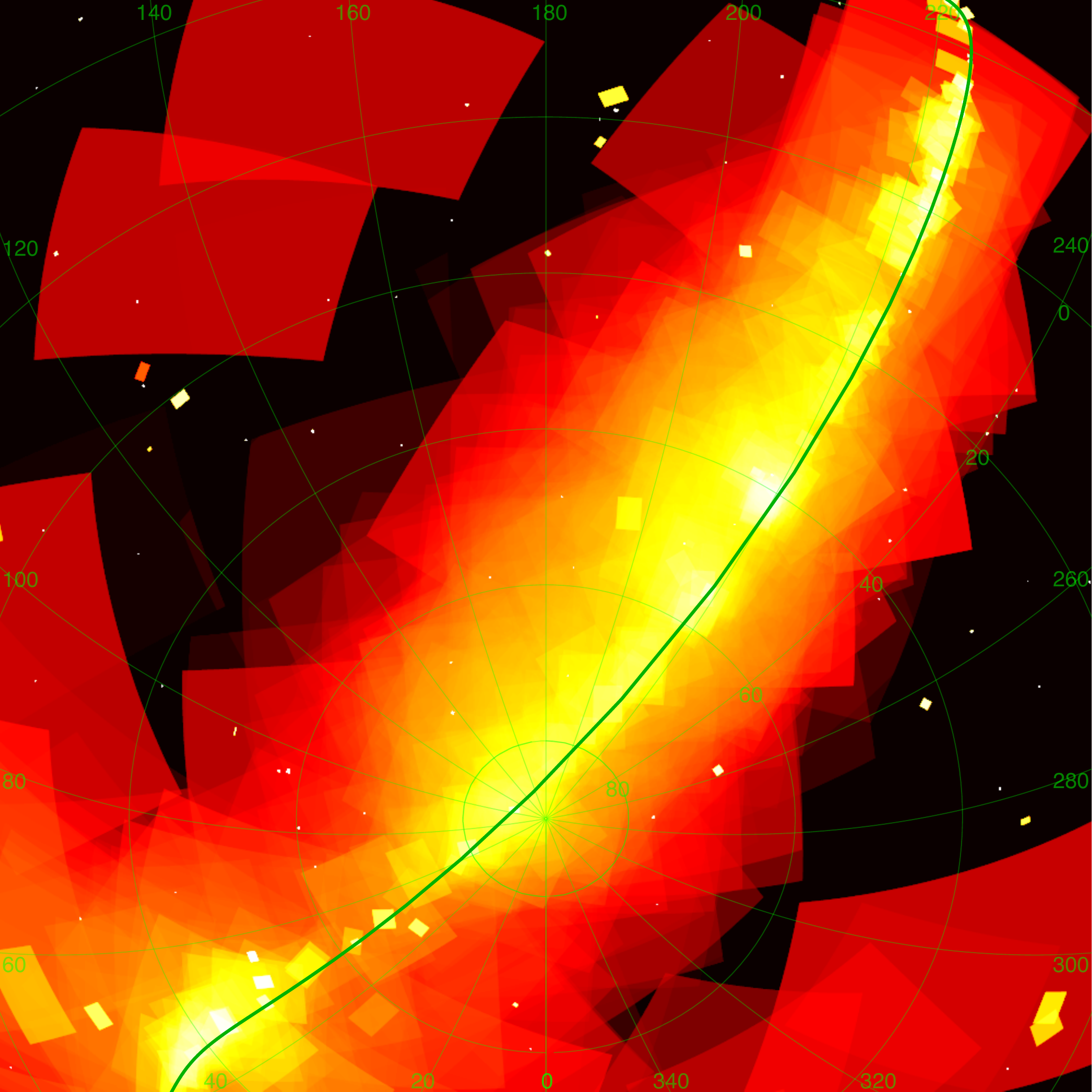}
\includegraphics[width=0.49\textwidth]{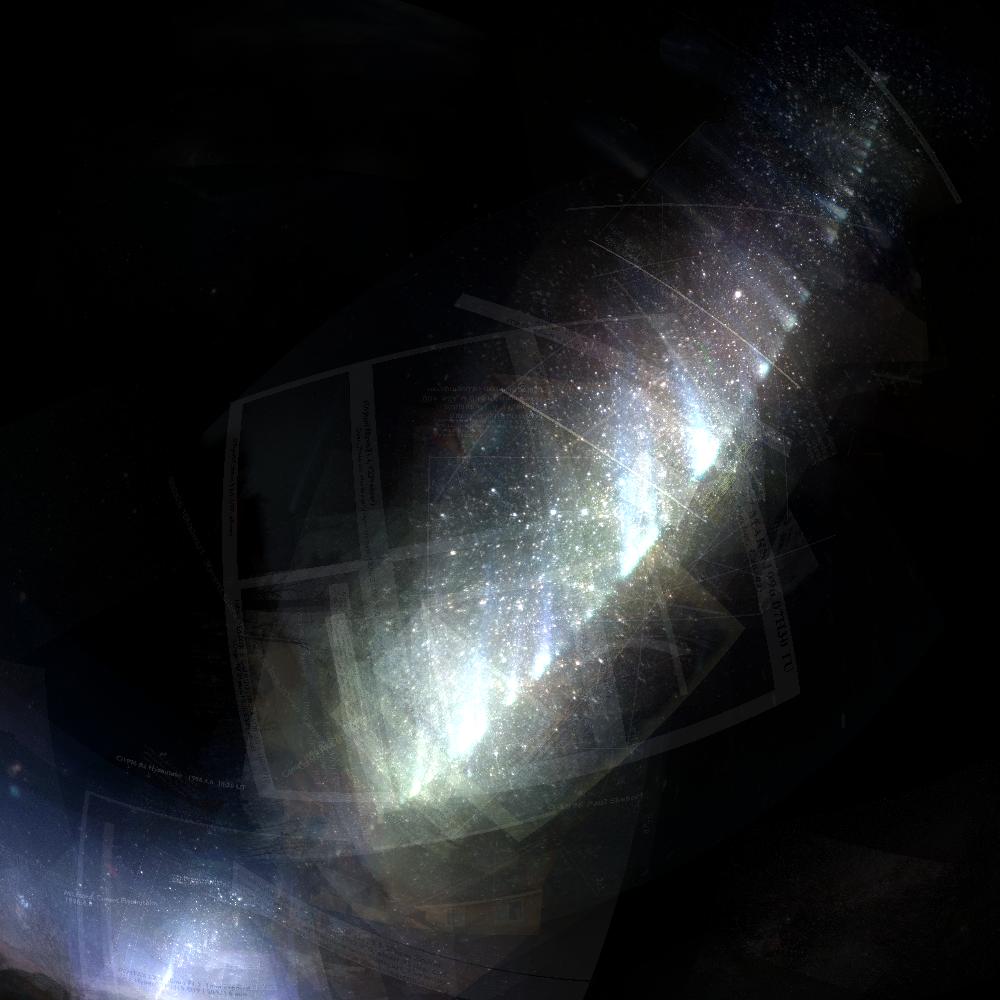}\hspace{1pt}%
\includegraphics[width=0.49\textwidth]{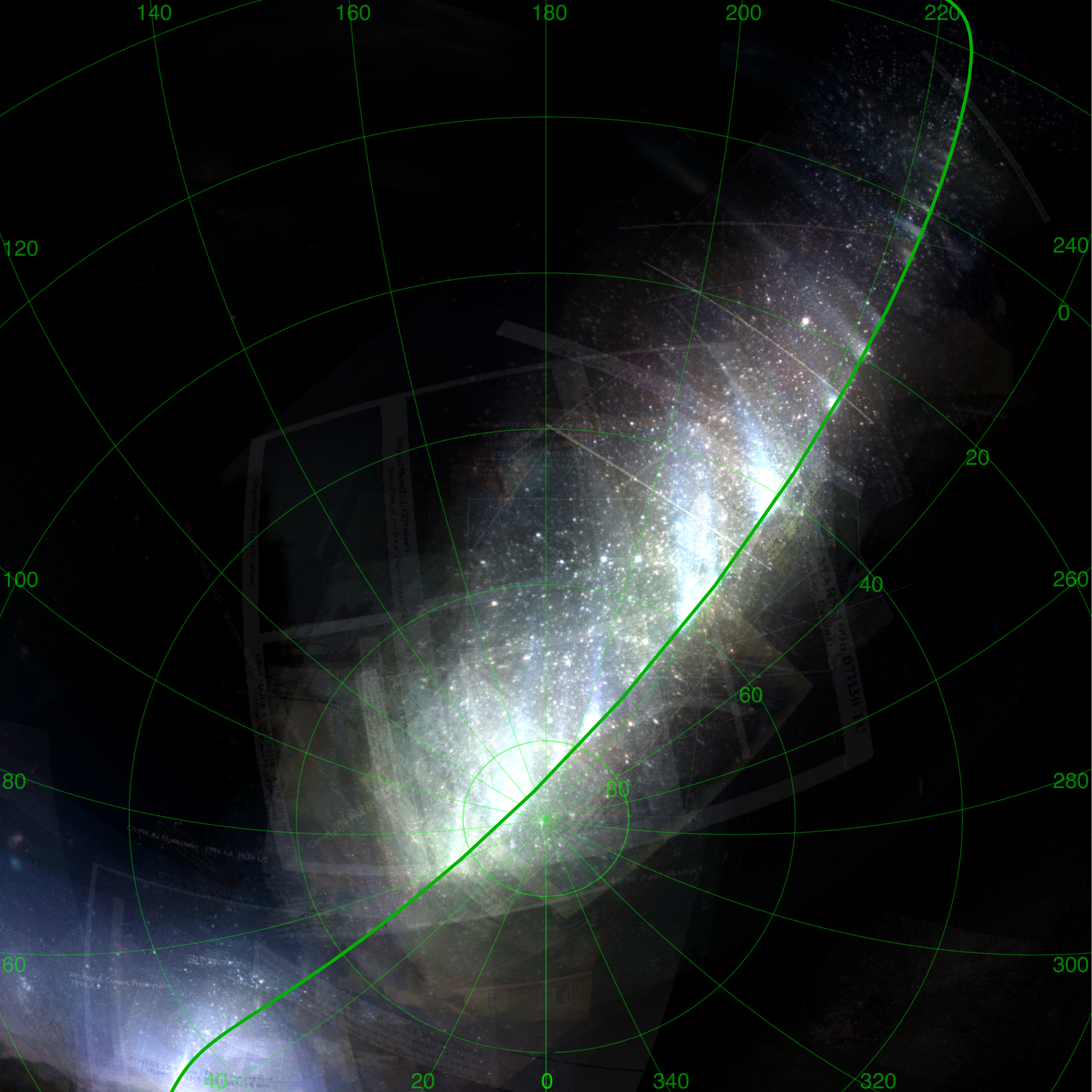}
\end{center}
\caption[hya]{Same as \figref{fig:footprints} but now for
  images found by a similar Web search for ``Comet Hyakutake''.  The
  search, performed on 2010~Oct~2, produced $1481$ JPEG images, of
  which $1019$ were recognized by \An\ as images of the night sky.
  \textsl{Top:} Pixel density map, with a log stretch.  The most
  heavily imaged areas appear in $152$ images.  The projection is
  zenithal equidistant (FITS WCS code ``ARC'', \citealt{wcsstandard}).
  \textsl{Bottom:} Co-added images.  The spectacular tail of the comet
  is clearly visible, as are many text labels, annotations, and image
  borders.  The right panels show the same images as the left panels
  but with a coordinate grid and the trajectory of the JPL ephemeris.
  \label{fig:hyakutake}}
\end{figure}


\end{document}